\documentclass{article}

\usepackage{PRIMEarxiv}
\usepackage[utf8]{inputenc} 
\usepackage[T1]{fontenc}    
\usepackage{hyperref}       
\usepackage{url}            
\usepackage{booktabs}       
\usepackage{amsfonts}       
\usepackage{dsfont}
\usepackage{nicefrac}       
\usepackage{microtype}      
\usepackage{lipsum}
\usepackage{fancyhdr}       
\usepackage{graphicx}       
\usepackage{float}
\usepackage{enumitem}
\usepackage[pdftex,dvipsnames]{xcolor}
\graphicspath{{media/}}     
\usepackage{longtable}
\usepackage{array}
\usepackage{multirow}
\usepackage{makecell}
\usepackage{amsmath}
\usepackage{xargs}
\usepackage{tabularx}
\usepackage{caption} 
\usepackage{subcaption}
\usepackage{csquotes}
\usepackage{xfrac}

\usepackage{algorithm}
\usepackage{algpseudocode}
\algrenewcommand\algorithmiccomment[1]{\hfill$\triangleright$~#1}

\renewcommand{\arraystretch}{1.10}
\newcolumntype{Y}{>{\raggedright\arraybackslash}X}
\newcolumntype{P}[1]{>{\raggedright\arraybackslash}p{#1}}
\newcolumntype{L}[1]{>{\raggedright\arraybackslash}p{#1}}
\newcolumntype{C}[1]{>{\centering\arraybackslash}p{#1}}

\usepackage[colorinlistoftodos,prependcaption,textsize=tiny]{todonotes}
\newcommandx{\unsure}[2][1=]{\todo[linecolor=red,backgroundcolor=red!25,bordercolor=red,#1]{#2}}
\newcommandx{\change}[2][1=]{\todo[linecolor=blue,backgroundcolor=blue!25,bordercolor=blue,#1]{#2}}
\newcommandx{\info}[2][1=]{\todo[linecolor=OliveGreen,backgroundcolor=OliveGreen!25,bordercolor=OliveGreen,#1]{#2}}
\newcommandx{\improvement}[2][1=]{\todo[linecolor=Plum,backgroundcolor=Plum!25,bordercolor=Plum,#1]{#2}}
\newcommandx{\thiswillnotshow}[2][1=]{\todo[disable,#1]{#2}}

\title{Procedural Generation of Conceptual 3D Ship Internal Arrangements for Design-Space Exploration
}

\author{
  K. Odendaal\\
  Delft University of Technology \\
  Delft, Netherlands\\
  \texttt{k.odendaal@tudelft.nl} \\
   \And
  R. de Winter \\
  Delft University of Technology \\
  Delft, Netherlands\\
  \texttt{r.dewinter@tudelft.nl} \\
}

\begin{document}
\maketitle

\vspace{-10pt}
\begin{abstract}
Ship internal arrangement is a complex spatial design problem where competing functions within a fixed volume affect spatial relationships, mass distribution, trim, and stability.
Exploring these trade-offs requires many layouts in a common format, but public general-arrangement drawings remain scarce, inconsistent, and predominantly two-dimensional.
Procedural generation can create controlled spatial alternatives, but broader design studies must also span different vessel families and changing requirements.
This study addresses these challenges by developing a common framework for generating conceptual 3D internal arrangements across six vessel families: bulk carriers, tankers, container vessels, offshore support vessels, patrol vessels, and yachts.
Each arrangement is represented as paired voxel grids and topological graphs, supporting spatial and connectivity analysis.
The framework is evaluated through three experiments: (i) comparison of 50,000 generated arrangements with 23 reconstructed real references, (ii) envelope-constrained reconstruction of one reference per family, and (iii) whether requested function volumes are reflected in generated layouts.
For each vessel family, the generated design space broadly aligns with real references within its sampling range.
Searching across alternative spatial subdivisions improves reconstruction across all families, with voxel accuracy reaching 75.5\%.
Fuel and ballast allocations respond near-proportionally to requested volume, while other functions depend on subdivision capacity, compartment eligibility, and discrete assignment.
Finally, a fixed-hull offshore support vessel application shows that arrangements with the same hull and requested functional programme produce different adjacency and trim outcomes, where five nondominated concepts emerge among 119 of 500 alternatives meeting the stability and programme criteria.
Together, the results support layout design-space exploration across vessel families and changing requirements, with spatial organisation remaining an important design variable beyond volume compliance.
The generator and conceptual general arrangement dataset are released publicly to support reproducible analysis, benchmarking, and future learning-based design studies.

\end{abstract}

\keywords{Ship internal arrangement \and Procedural generation \and Design-space exploration \and Data-driven design}

\section{Introduction}
\label{sec:introduction}

Ship internal arrangement is a highly complex spatial design problem~\cite{papanikolaouShipDesignMethodologies2014}.
Required ship functions must fit within a limited hull volume while maintaining suitable spatial relationships.
Their placement affects mass distribution and therefore influences draught, trim, and stability.
As a result, two arrangements can satisfy the same broad functional programme without being engineering-equivalent.
Early-stage design can therefore benefit from comparing realistic alternatives rather than selecting the first feasible arrangement.

However, systematic exploration requires enough layouts to reveal meaningful variation and a common representation for comparison.
General-arrangement drawings provide valuable evidence of real design practice, but public examples are limited and vary in detail.
Most are also two-dimensional, so converting them into comparable labelled 3D representations requires substantial reconstruction effort~\cite{haTwoStageConversionGA2025}.
Real design drawings provide individual realised solutions rather than controlled families of alternatives. They cannot naturally support systematic exploration of how an arrangement changes when one requirement varies while the others are held fixed.
Procedural generation provides a way to create controlled alternatives from explicit design inputs, allowing selected requirements to be varied while others remain fixed.
The motivation for this study is therefore to create and understand a sufficiently broad and controllable arrangement design space for systematic early-stage engineering comparison.

Computational methods already provide substantial capabilities for ship arrangement generation and optimisation.
Previous work includes packing methods~\cite{vanoersPackingApproachEarly2011}, network-based synthesis~\cite{gillespieGeneratingFunctionalComplexbased2013}, rule-based approaches~\cite{lepooleWARGEARRealTime2022}, and application-specific layout tools~\cite{kimReviewShipArrangement2024}.
More recent studies have also treated three-dimensional subdivision as a direct problem within arrangement design~\cite{deleonGenerativeAlgorithmsEarly2025}.
However, evidence remains limited on how the same generation procedure behaves across substantially different vessel families, including where its rules support useful variation and where they become restrictive.

To address this gap, this study develops a procedural generator for conceptual 3D internal arrangements across six vessel families: bulk carriers, tankers, container vessels, offshore support vessels, patrol vessels, and yachts.
The same generation process applies to every family, with hull geometry supplied as input and family-specific ranges defining dimensions and functional requirements.
The resulting dataset contains 50\,000 generated arrangements stored using paired graph and voxel representations.
The framework is evaluated by comparing it with reconstructed real arrangements, reconstructing selected reference arrangements, applying controlled changes to requested functional volumes, and exploring a fixed-hull design application.

The main contributions of this work are:
\begin{itemize}[topsep=1pt, itemsep=1pt]
\item A common generation approach that produces complete 3D internal arrangements for six unique vessel families.
\item A per-family analysis showing where the generated design space aligns with available real arrangements and where the procedure limits variation and control.
\item A fixed-hull demonstration showing that arrangements with similar functional requirements can differ in functional adjacency and trim because of their spatial organisation.
\item An openly released generator and dataset of 50\,000 paired graph and voxel arrangements for reuse in naval-architecture and data-driven research.
\end{itemize}

The generator scripts and resulting conceptual general arrangement dataset is released publicly~\cite{odendaal2026shipspacedataset,odendaal2026shipspacega}.

\section{Theoretical Background}
\label{sec:technical_background}

This section introduces the naval-architectural and graph-theoretic concepts
needed to interpret the analyses developed in the study. 

\subsection{Mass distribution, stability, and trim}
\label{sec:mass_stability_trim}

The coordinate system uses $x$ from aft to bow, $y$ from port to starboard, and $z$ upward from the keel.
For compartment masses $m_i$ at longitudinal positions $x_i$ and vertical positions $z_i$, the longitudinal and vertical centres of gravity are~\cite{papanikolaouShipDesignMethodologies2014,SchneekluthShipDesignEfficiency1998}
\begin{equation}
LCG=\frac{\sum_i m_i x_i}{\sum_i m_i},
\qquad
KG=\frac{\sum_i m_i z_i}{\sum_i m_i}.
\end{equation}
For comparison across vessel sizes, these quantities are reported as $LCG/L$ and $KG/D$, where $L$ and $D$ are ship length and depth.

Initial transverse stability follows the conventional metacentric relationships~\cite{papanikolaouShipDesignMethodologies2014,SchneekluthShipDesignEfficiency1998}
\begin{equation}
KM = KB + BM,
\qquad
GM = KM-KG,
\end{equation}
where $KB$ and $KG$ are the vertical centre of buoyancy and gravity, respectively.
$BM$ is the transverse metacentric radius, and $KM$ is the vertical position of the transverse metacentre.
Partly filled liquid tanks reduce the effective metacentric height through the free-surface effect:
\begin{equation}
GM_{\mathrm{corr}} = GM-\Delta GM_{\mathrm{FSM}},
\end{equation}
where $\Delta GM_{\mathrm{FSM}}$ is the free-surface correction.

At longitudinal equilibrium, the longitudinal centres of gravity and buoyancy align, but the vessel may still have non-zero trim~\cite{SchneekluthShipDesignEfficiency1998}.
Solved trim is defined from the equilibrium forward and aft draughts as
\begin{equation}
J_{\mathrm{trim}}=100\,\frac{|T_{\mathrm{aft}}-T_{\mathrm{fwd}}|}{L},
\end{equation}
where $J_{\mathrm{trim}}$ is the absolute trim expressed as a percentage of ship length.

\subsection{Functional topology and graph distance}
\label{sec:functional_topology}

A labelled voxel representation preserves the spatial extent and function of regions within the vessel.
A graph representation provides a complementary description of the relationships between those regions~\cite{gillespieNetworkScienceApproach2012,rothAnalysisGeneralArrangements2017,PawlingSeeingArrangements}.

The arrangement is represented on a regular three-dimensional grid of volume elements, or \emph{voxels}, with each voxel carrying one functional label.
A \emph{functional island} is a face-connected group of voxels with the same label.
Regions that touch only at an edge or corner therefore remain separate islands.

A Connected-Compartment Graph (CCGraph) contains one node for each functional island.
Two nodes are connected when their islands share a physical interface.
Internal void regions remain traversable, while cells outside the hull are excluded.
Figure~\ref {fig:six_family_representations} shows examples of the voxel and graph representations used in the study.

Distances through the CCGraph retain physical scale.
For adjacent nodes $u$ and $v$, let $\mathbf{c}_u$ and $\mathbf{c}_v$ denote their volume-weighted centroids, and let $\mathbf{c}_{uv}$ denote the area-weighted centroid of their shared interface.
The edge length follows the path from one region centroid to the shared interface and then to the second region centroid:
\begin{equation}
\ell_{uv}
=
\left\lVert \mathbf{c}_u-\mathbf{c}_{uv} \right\rVert_2
+
\left\lVert \mathbf{c}_v-\mathbf{c}_{uv} \right\rVert_2.
\end{equation}
The distance between two functional islands is the shortest weighted path through the CCGraph~\cite{dijkstraNoteTwoProblems1959}:
\begin{equation}
d_G(i,j)
=
\min_{P:i\rightarrow j}
\sum_{e\in P}\ell_e.
\end{equation}

\section{Related Work}
\label{sec:related_work}
This section positions the study across three related areas: computational generation of ship arrangements, engineering and graph-based arrangement evaluation, and using generated sample sets for design-space analysis.
Together, these areas motivate the need for a common representation and evaluation framework that can be applied consistently across multiple vessel families.

\subsection{Computational generation of ship arrangements}

Computer support for ship general-arrangement design predates modern optimisation methods.
Early systems such as the General Arrangement Design System formalised parts of the arrangement process as a computational workflow~\cite{carlsonGeneralArrangementDesign1987}.
Intelligent Ship Arrangements later separated space allocation to zone-decks from detailed placement~\cite{parsonsIntelligentShipArrangements2008}.
These approaches demonstrated that spatial and functional requirements could be explicitly represented during early arrangement design.

Later work expanded how arrangements are generated and explored.
Van Oers developed a packing-based method for generating feasible early-stage ship arrangements subject to design constraints~\cite{vanoersPackingApproachEarly2011}.
Duchateau extended this approach toward interactive exploration of alternative concepts and trade-offs~\cite{duchateauInteractiveEvolutionaryConcept2016}.
Optimisation methods have since been applied to a range of compartment and functional-allocation problems~\cite{kimReviewShipArrangement2024,adityaOptimisationNavalShip2021,wangApproachShipDeck2021,wangResearchIntelligentDesign2021,louvrosMultiobjectiveOptimisationEarly2022}.
More specialised systems include WARGEAR, which generates detailed surface-warship arrangements using probabilistic and network-based procedures~\cite{lepooleWARGEARRealTime2022}.

Recent work has moved toward explicit three-dimensional generation.
Padolecchia et al. developed a parametric workflow that links hull geometry and functional zoning to generating 3D general arrangements~\cite{padolecchia3DParametricModeling2025}.
The published case study focuses on a passenger-vessel main-deck problem and does not demonstrate that the same detailed parametrisation transfers directly to substantially different vessel architectures.
De León and Koelman present a stochastic BSP-based approach for producing three-dimensional hull subdivisions~\cite{deleonGenerativeAlgorithmsEarly2025}.
Subdivision itself is the design output rather than one stage within a complete labelled arrangement.
Together, these studies demonstrate substantial capability in computational arrangement generation, optimisation, and 3D representation.

\subsection{Arrangement evaluation and representation}

Ship arrangement assessment considers both vessel-level engineering behaviour and spatial relationships between ship functions.
Conventional naval-architecture measures describe mass location, trim, and stability~\cite{papanikolaouShipDesignMethodologies2014}.
Complementary spatial measures describe relationships such as adjacency and connectivity between functions~\cite{parsonsIntelligentShipArrangements2008,louvrosMultiobjectiveOptimisationEarly2022,klishinAvoidanceAdjacencyAssociation2021,gillespieGeneratingFunctionalComplexbased2013}.

Graph representations compactly describe these relationships.
Gillespie uses network models for early-stage arrangement reasoning~\cite{gillespieNetworkScienceApproach2012}, while Roth et al. convert generated arrangements into weighted graphs for comparison~\cite{rothAnalysisGeneralArrangements2017}.
Pawling and Andrews apply network descriptions to existing arrangements from different vessel domains and discuss both their value and limitations~\cite{PawlingSeeingArrangements}.
Related representations have supported routing and vulnerability studies~\cite{duchateauEarlyStageRouting2018,vandiessenIntegratingVulnerabilityAnalysis2022}.
More recent work has used networks to describe relationships within three-dimensional general arrangements~\cite{dowlingCharacterizingThreedimensionalGeneral2024}.
These studies establish physical and graph-based descriptions as complementary ways to compare internal arrangements.

\subsection{Design space analysis}

Large generated sample sets have also been used to study engineering design spaces and support later data-driven design~\cite{hongDeepJEB3DDeep2024}.
Although focused on external hull geometry rather than internal arrangements, Ship-D provides a relevant maritime example.
It defines a common parametric hull representation, generates a large set of hull designs, tests that representation by reconstructing real hulls, and demonstrates its use in downstream engineering optimisation~\cite{bagazinskiShipDShipHull2023a}.
Jaspers analysed a sample set generated with the Delft packing approach and found that part of its apparent structure was linked to the separate generation runs used to create it~\cite{noneElucidatingFamiliesShip2017}.

High-quality internal-arrangement data remain limited~\cite{peiReviewResearchShip2026}.
Available general-arrangement drawings provide real-world cases for comparison, but converting them into consistent, labelled 3D arrangements requires substantial reconstruction~\cite{haTwoStageConversionGA2025}.
Generated sample sets offer scale and controlled variation, whereas reconstructed real arrangements anchor comparison to existing designs.

Although some methods have been applied to more than one vessel~\cite{vanoersPackingApproachEarly2011,duchateauInteractiveEvolutionaryConcept2016}, published evaluations commonly focus on individual vessel families or specific design problems.
As a result, there is limited information on how the same generation process behaves across different vessel types. 
It remains unclear where the same rules support useful variation and requirement control, and where they become restrictive.

\section{Methodology}
\label{sec:methodology}

This study uses one fixed procedural generator configuration to construct conceptual 3D internal arrangements across six vessel families.
Each sample begins with a vessel family, principal dimensions, and a compatible hull geometry.
Further inputs control the spatial subdivision, requested functional programme, and target longitudinal and vertical mass positions.
The hull geometry is voxelised to define a physically scaled $64\times32\times24$ occupancy mask.
The internal arrangement is represented on the same grid so that volumes, distances, and mass properties retain their physical meaning.

\begin{figure}[t]
    \centering
    \includegraphics[width=\linewidth]{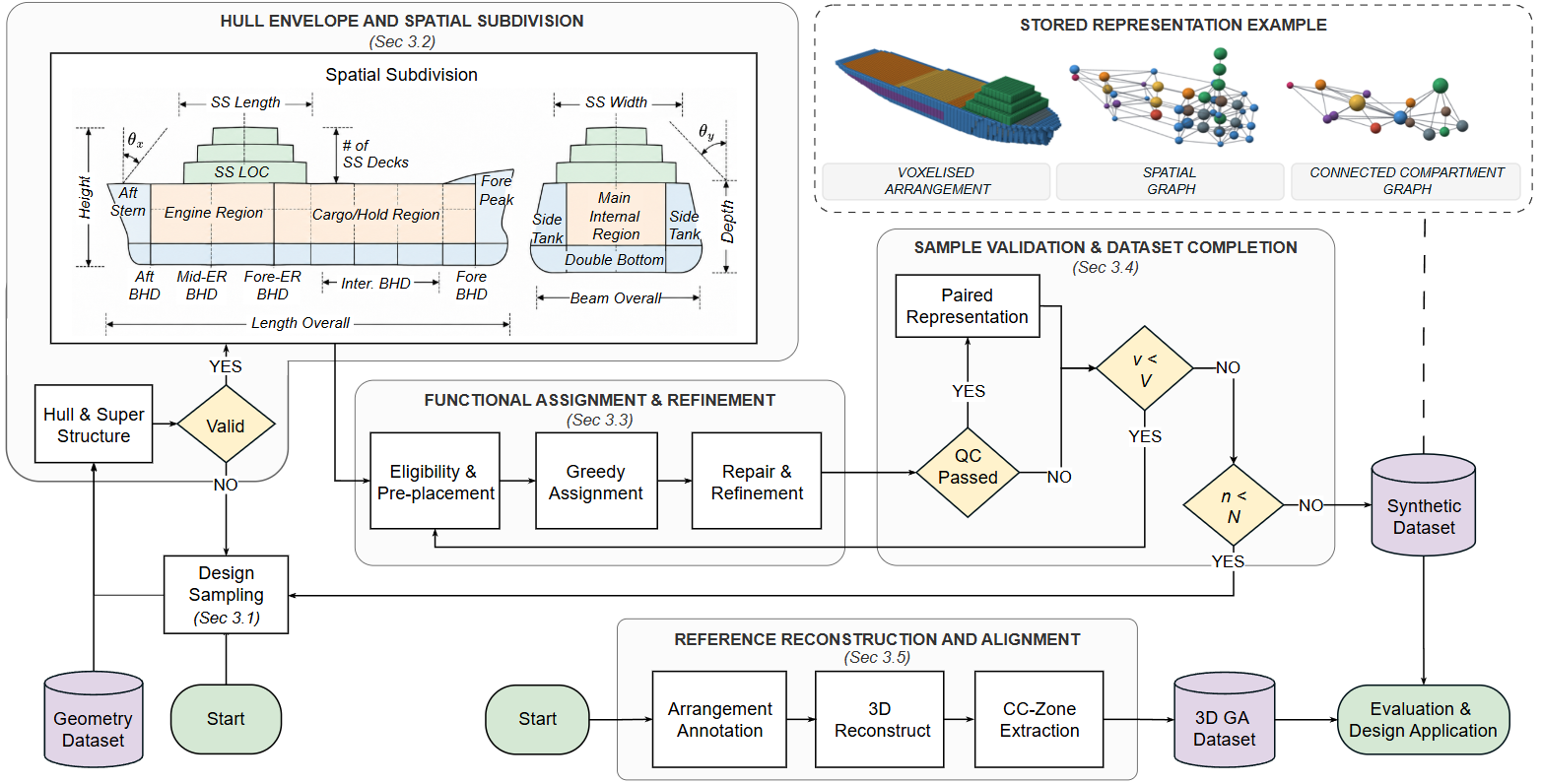}
    \caption{Overview of the procedural generation and evaluation workflow.
    A sampled design intent and hull geometry define the spatial subdivision, followed by functional assignment, correction and refinement, quality control, and representation export.
    Nested loops drive iteration: an inner variant cycle generates multiple unique functional layouts per subdivision (\textit{V}), and an outer loop controls the total sample count (\textit{N}).
    A parallel reconstruction branch maps real general arrangements to the same representations for external comparison.}
    \label{fig:methodology_workflow}
    \vspace{-10pt}
\end{figure}

Figure~\ref{fig:methodology_workflow} summarises the procedural generation stages.
The occupied hull volume is first divided into an unlabelled spatial subdivision.
Functional labels are then assigned under programme, spatial, and mass-position constraints.
Repair and refinement improve the assignment without resampling the hull or subdivision parameters.
Accepted arrangements are stored as paired graph and voxel representations of the same design.
The process is stochastic across seeds but reproducible for a fixed seed and configuration.

Construction rules, data quality control (QC), and external validation serve different purposes.
Construction rules define the arrangements that can be generated.
QC checks whether a sample is internally consistent with those rules before it enters the generated sample set.
External validation examines how the resulting design space relates to reconstructed real arrangements and responds to controlled design changes, as described in Section~\ref{sec:data_experimental_design}.
Passing the QC therefore establishes consistency and validity with the generator rules.

\subsection{Design sampling and parameterisation}
\label{sec:procedural_generation}

Each arrangement begins from a sampled design intent.
The six vessel families are bulk carriers, tankers, container vessels, offshore support vessels, patrol vessels, and yachts.
Principal dimensions $L$, $B$, and $D$ are sampled from family-specific ranges together with a compatible hull geometry.
Further controls define the superstructure, spatial subdivision, requested functional programme, and target $LCG/L$ and $KG/D$.
Section~\ref{sec:data_experimental_design} gives the sampling ranges used in the reported design-space analysis.

The requested functional programme contains seven volume classes: engine room, machinery, cargo or mission space, stores, accommodation, fuel, and ballast.
Steering gear is a mandatory function, not a sampled volume request.
Void represents residual capacity and is not requested directly.
The sampled $LCG/L$ and $KG/D$ targets guide functional placement.

All vessels use the same $64\times32\times24$ main-hull grid in index space.
Physical cell dimensions vary with the sampled dimensions:
\begin{equation}
\Delta x=\sfrac{L}{64},
\qquad
\Delta y=\sfrac{B}{32},
\qquad
\Delta z=\sfrac{D}{24}.
\end{equation}
The resolution was selected during the development of the generator as a standard discretisation that preserves key internal features, such as double bottoms and side-tank bands, across various vessel sizes.
The reconstructed general arrangements provided insights into the range of dimensions.
This approach ensures a consistent representation while maintaining the vessel-scale volumes, distances, centroids, and mass properties.
The remaining sampled controls define the subdivision geometry.
They specify the peak and engine-region boundaries, vertical and transverse divisions, and superstructure geometry.

\subsection{Hull envelope and spatial subdivision}

The hull geometry is not generated procedurally.
A family-compatible reference stereolithography (STL) hull is selected from the hull library, aligned, and scaled independently in the longitudinal, transverse, and vertical directions to match the sampled $L$, $B$, and $D$.
It is then voxelised on the physical grid.
The resulting hull mask defines the available internal volume.
A sampled superstructure is added above the local deck profile.

The production hull library contains 17 hull forms: five transport hulls, six work-vessel hulls, three yacht hulls, and three patrol hulls.
The transport set is shared by the bulker, tanker, and container-vessel families, while the remaining sets provide the hulls for OSV, yacht, and patrol-vessel generation.
For each generated sample, a hull is randomly selected from the compatible hull pool for the corresponding vessel family to introduce variation in the external envelope.
Appendix~\ref{app:stl_provenance} documents detailed hull geometries and provenance.

Before functional assignment, the occupied envelope is divided into an unlabelled spatial subdivision.
Longitudinal boundaries define the aft peak, engine region, main body, and fore peak, with additional divisions through the main body.
Vertical boundaries define the double bottom and lower and upper tiers.
Selected regions are divided transversely into port, centre, and starboard zones.
The superstructure is partitioned separately.
Every zone is clipped to the occupied hull or superstructure mask, so its usable volume follows the physical envelope.
Here, $N_h$ denotes the number of main-hull subdivision zones.

Subdivision and functional assignment are separated so that labels are assigned to coherent spatial zones rather than directly to individual voxels.
This keeps the spatial boundaries explicit and limits fragmented functional regions during assignment.

Adjacent zones form the generator-native subdivision graph.
Port and starboard zones are also linked as mirror partners where symmetric assignment is required.
Each zone type is associated with a permitted set of functions, representing early-stage engineering practice for subdivision, machinery spaces, tank location, and symmetry.
Table~\ref{tab:zone_eligibility} summarises these eligibility rules.

\begin{table}[t]
\centering
\scriptsize

\begin{subtable}[t]{0.60\linewidth}
\centering
\setlength{\tabcolsep}{3pt}
\renewcommand{\arraystretch}{1.2}

\begin{tabular}{p{0.20\linewidth} p{0.73\linewidth}}
\hline
\textbf{Subdivision zone} & \textbf{Eligible function compartments} \\
\hline
\text{Aft Peak}
& \text{Ballast}, \text{Steering gear}, \text{Void} \\

\text{Engine Region}
& \text{Engine room}, \text{Fuel}, \text{Machinery}, \text{Void} \\

\text{Double Bottom}
& \text{Ballast}, \text{Fuel}, \text{Void} \\

\text{Main Lower}
& \text{Ballast}, \text{Cargo/Mission}, \text{Fuel},
  \text{Machinery}, \text{Stores}, \text{Void} \\

\text{Main Upper}
& \text{Accommodation}, \text{Cargo/Mission},
  \text{Machinery}, \text{Stores}, \text{Void} \\

\text{Fore Peak}
& \text{Ballast}, \text{Machinery}, \text{Void} \\

\text{Superstructure}
& \text{Accommodation}, \text{Void} \\

\text{Engine Upper}
& \text{Machinery}, \text{Void}, and either
  \text{Cargo/Mission} or \text{Accommodation/Stores}$^\dagger$ \\

\text{Side/wing Region}
& \text{Ballast}, \text{Fuel}, \text{Void} \\
\hline
\end{tabular}

\vspace{2pt}
\raggedright
{\scriptsize $^\dagger$ Vessel-family dependency.}

\caption{Functional eligibility of the main subdivision zones. 
Void is permitted throughout as residual capacity. 
Fuel eligibility in side zones is further restricted by vessel family and longitudinal position.}
\label{tab:zone_eligibility}
\end{subtable}%
\hfill%
\begin{subtable}[t]{0.35\linewidth}
\centering
\setlength{\tabcolsep}{2pt}
\renewcommand{\arraystretch}{1.2}

\begin{tabular}{@{}l r@{}}
\hline
\textbf{Function} & 
\textbf{$\text{Relative mass intensity: } \tilde{\rho}$ (t/m$^3$)} \\
\hline
\text{Engine room}     & 5.00  \\
\text{Steering gear}   & 3.50  \\
\text{Machinery}       & 2.50  \\
\text{Cargo/Mission}$^\dagger$   & 2.00  \\
\text{Stores}          & 1.50  \\
\text{Accommodation}   & 0.50  \\
\text{Void}            & 0.05  \\
\hline
\textbf{Liquid} & 
\textbf{$\text{Physical density: } \rho$ (t/m$^3$)} \\
\hline
\text{Fuel}            & 0.85  \\
\text{Ballast}         & 1.025 \\
\hline
\end{tabular}

\vspace{2pt}
\raggedright
{\scriptsize $^\dagger$ Yachts Cargo/Mission takes on accommodation
value of $0.50$.}

\caption{Functional mass parameters used in the common mass model.
Fuel and ballast use representative physical liquid densities. 
The remaining entries are relative mass-intensities.}
\label{tab:mass_proxies}
\end{subtable}
\vspace{-7pt}
\caption{Subdivision eligibility and functional mass-property parameters used by the generator.}
\label{tab:methodology_parameters}
\vspace{-10pt}
\end{table}

The realised subdivision fixes the zone geometry, adjacency, eligible capacity, and spatial granularity while leaving the functional labels undetermined.
Multiple functional assignments can therefore be produced from the same subdivision, while varying the subdivision introduces additional spatial variation.

\subsection{Functional assignment and refinement}
\label{sec:programme_assignment}

A requested programme may not fit a realised subdivision because some functions have limited eligible capacity.
The procedural generator therefore distinguishes three programme budget states ($\mathbf{b}$):
\begin{equation}
\mathbf{b}^{\mathrm{req}}
\rightarrow
\mathbf{b}^{\mathrm{eff}}
\rightarrow
\mathbf{b}^{\mathrm{ach}},
\end{equation}
where $\mathbf{b}^{\mathrm{req}}$ is the sampled request, $\mathbf{b}^{\mathrm{eff}}$ accounts for the eligible capacity of the realised subdivision, and $\mathbf{b}^{\mathrm{ach}}$ is the final realised volume distribution.
Requests that exceed reachable capacity are reduced, and the excess programme fraction is reassigned to functions with remaining eligible capacity where possible.
Residual capacity can remain as void, providing slack for discrete zone assignment.
A candidate is rejected if a feasible effective programme cannot be formed.

Functional assignment begins after this capacity check.
Mandatory placements first establish functions with strong positional constraints.
Steering gear is placed in the aft peak.
The engine room is seeded low in the engine region and expanded through adjacent eligible zones.
Family-specific accommodation and stores rules are applied where required.
Mirror partners are assigned together when symmetry is required.

Following pre-placement, the remaining zones are assigned using a seed-perturbed greedy assignment procedure.
Zones are processed sequentially in a deterministic order, and at each step the highest-scoring eligible function ($q$) is selected according to
\begin{equation}
S_{iq}
=
D_q
F_{iq}
P^{\mathrm{space}}_{iq}
P^{\mathrm{mass}}_{iq}
(1+\epsilon_{iq}),
\end{equation}
where $D_q$ is the remaining programme deficit and $F_{iq}$ reduces excessive overshoot when a zone is much larger than that deficit.
$P^{\mathrm{space}}_{iq}$ represents a bounded multiplier that favours function-specific longitudinal, vertical, and zone-type locations among the zones eligible for function $q$.
$P^{\mathrm{mass}}_{iq}$ favours placements that move the running $LCG/L$ and $KG/D$ toward their sampled targets.
The term $\epsilon_{iq}$ introduces a small seed-controlled perturbation.

Eligibility and mandatory pre-placement are applied before this score.
The spatial preferences encode broad family-dependent tendencies, such as machinery near the engine region, accommodation in upper regions, and ballast in double-bottom and lower regions.
Appendix~\ref{app:compartment_assignment} gives the numerical spatial-preference functions.
Programme deficits and running mass moments are updated after each assignment.

The procedure is a constrained stochastic assignment heuristic, not a global arrangement optimiser.
Different seeds can therefore produce different functional assignments from the same subdivision and programme.
Retaining the requested, effective, and achieved programmes separately allows later analysis to separate requested design changes from limits imposed by eligible capacity and discrete assignment.

Sequential assignment does not revisit earlier decisions, so a bounded repair stage corrects known inconsistencies without restarting the arrangement.
Repairs reduce avoidable void, rebalance large programme errors, restore engine-room contiguity, and force mirrored assignments in case of accidental mismatch.
Fuel assignments are also corrected where they violate the encoded separation rules.
Centreline fuel is not permitted directly above the engine room, and side fuel is not permitted abreast central accommodation.

Unused superstructure regions assigned void are removed from the represented envelope.
This is the sole geometric exception after the spatial subdivision has been constructed.

After repair, a constrained label refinement improves agreement with the sampled mass-position targets without changing the achieved functional programme.
The refinement evaluates coordinated relabelling among eligible centre zones or mirrored zone groups.
Candidate changes are accepted only when they reduce $LCG/L$ and $KG/D$ tracking error while preserving the achieved functional-volume totals.
Eligibility, engine-room contiguity, mirror consistency, fuel-separation rules, and residual-space limits remain fixed.
If no feasible improving change is found, the repaired assignment is retained. The refinement is restricted to functional relabelling, with the hull geometry and spatial subdivision held fixed.

\subsection{Quality control and paired representation}
\label{sec:builder_qc_outputs}

A completed arrangement enters the generated dataset only after passing the active quality-control criteria in Table~\ref{tab:builder_qc}.
Each criterion reflects a regulatory design principle, conventional arrangement practice, or a generator-specific computational requirement. 

\begin{table}[t]
\centering
\scriptsize
\renewcommand{\arraystretch}{1.15}
\begin{tabular}{p{0.22\linewidth} p{0.39\linewidth} p{0.33\linewidth}}
\hline
\textbf{Check group} &
\textbf{Active acceptance condition} &
\textbf{Engineering basis} \\
\hline
Subdivision and assignment
&
Non-empty zones cover the occupied volume; every zone receives one eligible label; required engine-room and accommodation regions are present.
&
Protected double-bottom and peak zoning is informed by SOLAS II-1/9 and II-1/12; MARPOL Annex I/19 provides the closest analogue for tanker protective spaces \cite{IMO_SOLAS2024,IMO_MARPOL2022}. Other eligibility rules follow early-stage arrangement practice \cite{SchneekluthShipDesignEfficiency1998,papanikolaouShipDesignMethodologies2014}.
\\
Spatial coherence
&
The engine room forms one connected region; steering gear remains in the aft peak; paired port/starboard zones retain matching labels.
&
Arrangement practice \cite{SchneekluthShipDesignEfficiency1998,papanikolaouShipDesignMethodologies2014}; the exact connectivity and mirror-matching conditions are generator constraints.
\\
Programme and residual volume
&
The effective programme is capacity-feasible; each functional-volume error is at most $\max(0.12,\,2/N_h)$; interior void is at most its sampled target plus $\max(0.05,\,2/N_h)$; wing and double-bottom void is limited to $0.12$ of occupied hull volume.
&
Generator-specific acceptance tolerances for discrete zone assignment.
The $2/N_h$ term increases the tolerance for subdivisions containing fewer zones, where a single zone represents a larger fraction of the arrangement.
The fixed limits are empirical generator thresholds instead of regulatory criteria.
\\
Fuel separation
&
Fuel is excluded from the forepeak; centreline fuel is not placed directly above the engine room; side fuel is not placed abreast central accommodation.
&
MARPOL Annex I/16 provides the basis for forepeak oil exclusion \cite{IMO_MARPOL2022}. SOLAS II-2/4 provides the broader fire-safety basis for fuel-oil arrangements \cite{IMO_SOLAS2024}; the additional adjacency restrictions are conservative design proxies.
\\
\hline
\end{tabular}
\caption{Active quality-control criteria and their engineering basis.
Regulatory references indicate the closest underlying design principles.}
\label{tab:builder_qc}
\vspace{-10pt}
\end{table}

The programme-error tolerance is defined as $\max(0.12,\,2/N_h)$ to accommodate error introduced by the discrete spatial subdivision.
The $2/N_h$ term scales the tolerance with subdivision granularity and corresponds approximately to two mean zone shares if the zones were equal in size.
The fixed 0.12 lower bound was introduced empirically during generator development to reject large programme mismatches without excessive sample rejection as the number of subdivision zones increases.
The tolerance is therefore a generator-specific construction criterion rather than a regulatory or theoretically derived limit. 
Emergent mass properties, including $LCG$ and $KG$, are recorded as diagnostics rather than QC rejection criteria.
Stability is evaluated separately in the fixed-hull engineering application and is not used to determine whether a generated arrangement passes QC.



Each QC-accepted arrangement is exported as both a paired graph and a voxel representation.
The exported graph is the generator-native subdivision graph, with one node for each spatial subdivision zone and edges representing physical adjacency.
Each graph node also carries the functional label assigned to its corresponding zone.
The voxel representation inherits these labels on the physical grid, with a direct mapping between subdivision zones and voxels.

The subdivision graph and CCGraph describe different aspects of an arrangement.
The subdivision graph records the spatial zones created during generation, whereas the CCGraph describes the connected functional regions of the completed labelled arrangement.
All graph-based arrangement results in this study use the CCGraph defined in Section~\ref{sec:functional_topology}.

Dataset construction uses two nested sampling loops.
For each sampled hull, design intent, and spatial subdivision, up to $V=5$ seed-varied functional assignments are generated while the subdivision is held fixed.
Only variants that pass QC are retained so that a subdivision may contribute between one and five accepted arrangements.
A new subdivision is then sampled and the process repeats until the per-family sample target is reached.
After processing the $V$ variants, a new subdivision is sampled and the process repeats until the target dataset size of $N=50\,000$ accepted arrangements is reached. 
This sampling strategy introduces variation through both the spatial subdivision and the functional assignment, while also providing multiple functional layouts for some shared subdivisions.

\subsection{Reference reconstruction and representation alignment}
\label{sec:real_synthetic_abstraction}

Generated and real arrangements must be represented at the same level of abstraction for comparison.
Available general-arrangement drawings are therefore reconstructed as labelled three-dimensional volumes and mapped to the same functional classes used by the generator.
This removes drawing-specific naming differences while retaining the spatial information needed for comparison.
Reconstruction involves interpretation because the source drawings vary in annotation, geometry, and level of detail~\cite{haTwoStageConversionGA2025}.

The same analysis is then applied to both reconstructed and generated arrangements.
Functional volumes and locations are calculated from the labelled physical volumes.
Arrangement mass properties use the common model in Section~\ref{sec:engineering_evaluation}.
Connected regions with the same functional label are converted to the CCGraph defined in Section~\ref{sec:functional_topology}.
The two sources can therefore be compared through common measures of functional composition, spatial location, functional topology, and mass distribution without requiring their original subdivision structures to match.

In this study, the real arrangements are reference designs rather than an independent statistical sample of the fleet.
They also informed development of the generator and its family-specific parameter ranges.
The comparisons therefore provide plausibility anchors for the generated design space.

\section{Data and Experimental Design}
\label{sec:data_experimental_design}

The evaluation combines a generated multi-family sample set with reconstructed real arrangements used as reference cases.
This section defines the two data sources, the evaluation measures, and the experimental design.

\subsection{Generated and reference arrangements}
\label{sec:generated_reference_arrangements}

The primary generated dataset contains 50\,000 QC-accepted arrangements, balanced across the six vessel families. 
All arrangements are produced using the procedure described in Section~\ref{sec:methodology}.
Table~\ref{tab:sampling_ranges} summarises the family-specific sampling ranges used to construct the design space. 
Principal dimensions define vessel scale, while the remaining parameters control the spatial subdivision, superstructure, requested functional programme, and mass-position targets.

\begin{table}[t]
\centering
\scriptsize
\renewcommand{\arraystretch}{1.2}
\setlength{\tabcolsep}{3.2pt}
\begin{tabular}{lccccc}
\hline
\textbf{Family} &
$\boldsymbol{L}$ &
$\boldsymbol{B/L}$ &
$\boldsymbol{D/L}$ &
$\boldsymbol{LCG/L}$ &
$\boldsymbol{KG/D}$ \\
\hline
Bulker
& 150--280 & .14--.19 & .058--.081 & .42--.52 & .45--.62 \\
Tanker
& 75--150 & .14--.18 & .080--.095 & .42--.52 & .45--.62 \\
Container vessel
& 230--300 & .12--.18 & .068--.082 & .42--.53 & .46--.64 \\
OSV
& 50--95 & .18--.27 & .084--.113 & .40--.55 & .44--.62 \\
Patrol vessel
& 55--85 & .12--.15 & .115--.145 & .40--.52 & .46--.64 \\
Yacht
& 50--100 & .13--.22 & .095--.155 & .42--.56 & .46--.63 \\
\hline
\end{tabular}

\vspace{3pt}

\textbf{(a) Principal dimensions and mass-position targets}

\vspace{6pt}

\begin{tabular}{lccccccc}
\hline
\textbf{Family} &
\textbf{ER fwd.} &
\textbf{Main div.} &
\textbf{ER share} &
\textbf{SS $x/L$} &
\textbf{SS $L/L$} &
\textbf{SS $B/B$} &
\textbf{SS decks} \\
\hline
Bulker
& .22--.34 & 5--9  & .60--.75 & .05--.15 & .08--.15 & .70--.90 & 4--6 \\
Tanker
& .22--.34 & 6--12 & .60--.75 & .05--.15 & .08--.15 & .70--.90 & 3--4 \\
Container vessel
& .22--.36 & 3--5  & .55--.70 & .15--.30 & .05--.08 & .50--.70 & 7--10 \\
OSV
& .58--.72 & 2--4  & .40--.55 & .75--.90 & .15--.28 & .75--.95 & 2--3 \\
Patrol vessel
& .55       & 3--6  & .35--.50 & .15--.40 & .35--.60 & .65--.85 & 1--2 \\
Yacht
& .50       & 2--4  & .50--.65 & .40--.70 & .35--.60 & .80--.95 & 2--3 \\
\hline
\end{tabular}

\vspace{3pt}

\textbf{(b) Longitudinal subdivision and superstructure}

\vspace{6pt}

\begin{tabular}{lcccccccc}
\hline
\textbf{Family} &
\textbf{Engine} &
\textbf{Mach.} &
\textbf{Cargo/mission} &
\textbf{Stores} &
\textbf{Accom.} &
\textbf{Fuel} &
\textbf{Ballast} &
\textbf{Prog. cap} \\
\hline
Bulker
& .05--.08 & .008--.020 & .53--.63 & .008--.018 & .04--.07 & .04--.08  & .13--.20 & .92--.98 \\
Tanker
& .05--.08 & .02--.06   & .45--.58 & .02--.04   & .03--.10 & .035--.07 & .11--.20 & .92--.98 \\
Container vessel
& .03--.07 & .015--.035 & .55--.70 & .005--.025 & .03--.06 & .04--.08  & .08--.14 & .90--.98 \\
OSV
& .04--.08 & .06--.10   & .16--.42 & .02--.07   & .18--.28 & .08--.16  & .10--.19 & .88--.96 \\
Patrol vessel
& .08--.12 & .14--.24   & .03--.08 & .10--.15   & .28--.34 & .03--.07  & .04--.08 & .90--.98 \\
Yacht
& .045--.085 & .08--.14 & .12--.22 & .04--.10   & .28--.42 & .02--.06  & .04--.09 & .88--.95 \\
\hline
\end{tabular}

\vspace{3pt}

\textbf{(c) Requested functional programme}

\vspace{4pt}

\parbox{0.98\linewidth}{\scriptsize
\textit{Notes.}
ER = engine room; SS = superstructure.
ER fwd. denotes the forward boundary of the engine region, measured from aft as a fraction of $L$.
Main div. is the sampled number of longitudinal main-body divisions.
ER share is the fraction of the engine region assigned to its aft engine-room portion before functional assignment.
Programme cap is the maximum combined requested functional fraction; remaining assignable volume may be left as void.
$LCG/L$ and $KG/D$ are placement targets rather than acceptance limits.}
\caption{Family-specific sampling ranges used to construct the generated design space. $L$ is given in metres; geometric ratios and positions are dimensionless, while Main div. and SS decks are counts. Functional-programme entries are requested fractions of assignable internal volume.}
\label{tab:sampling_ranges}
\vspace{-10pt}
\end{table}

\begin{figure}[t]
\centering
\includegraphics[width=0.95\linewidth]{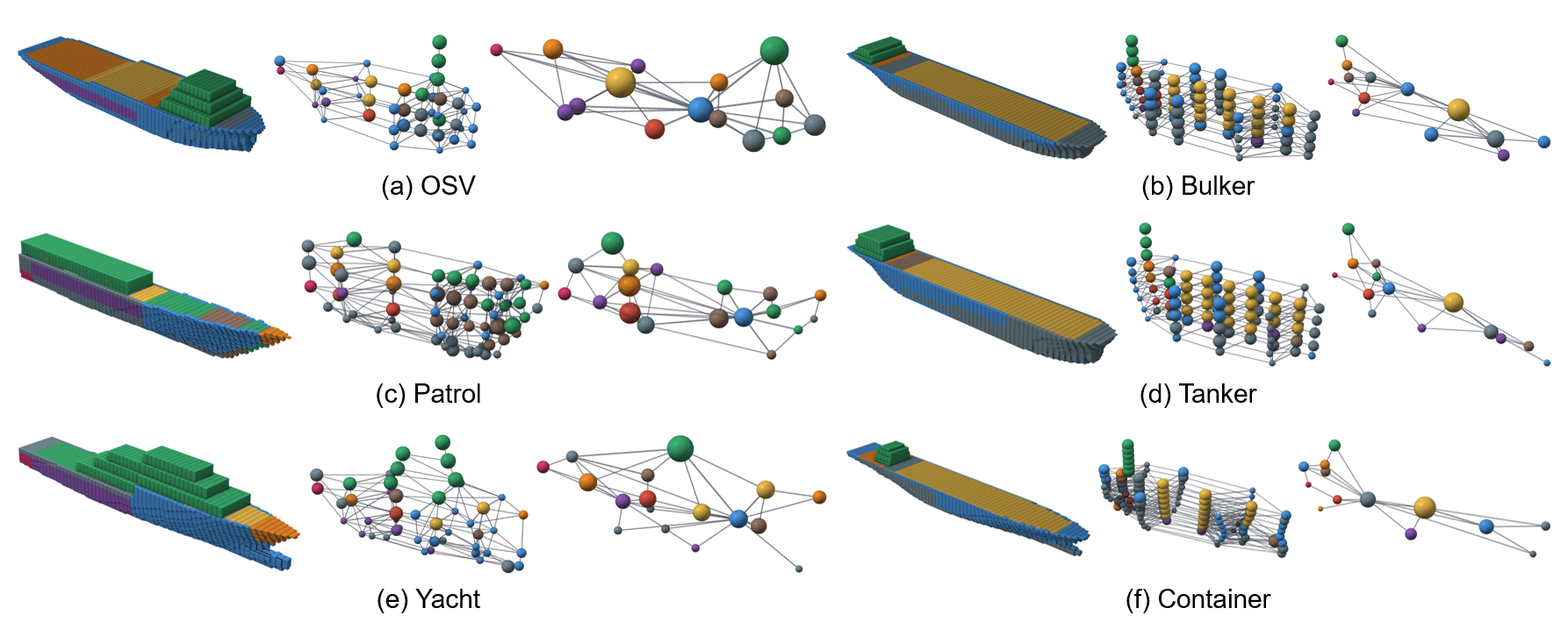}
\caption{Example representations across the six vessel families: (left) labelled voxel arrangement, (middle) generator-native subdivision graph, and (right) derived CCGraph. The examples illustrate the common representations used in the study rather than statistical representatives of the fleet.}
\label{fig:six_family_representations}
\vspace{-10pt}
\end{figure}

Several geometric rules are common to all vessel families. The aft-peak boundary is sampled between $0.04L$ and $0.08L$, and the fore-peak boundary between $0.90L$ and $0.95L$. 
Double-bottom height is set to $0.06B$, bounded between 1.0 and 2.5~m. 
The nominal side-region width on each side is also based on $0.06B$, bounded between 0.76 and 2.0~m.
Selected family-specific ranges were informed by the reconstructed real arrangements where suitable reference information was available. 
The peak and double-bottom parameterisations are standards-informed conceptual design proxies, consistent with the engineering basis in Table~\ref{tab:builder_qc}. 
The side-region width is a generator-specific geometric rule.
Figure~\ref{fig:six_family_representations} shows example arrangements from all six vessel families using the labelled voxel, generator-native subdivision-graph, and derived CCGraph representations.

The real-reference set contains 23 reconstructed general arrangements aligned to the common representation described in Section~\ref{sec:real_synthetic_abstraction}. 
It comprises six yachts, five OSVs, and three each of bulkers, tankers, container vessels, and patrol vessels. 
These arrangements provide reference cases for the evaluations described below. 
Their source documentation is provided in Appendix~\ref{app:real_ga_sources}.

\subsection{Engineering evaluation measures}
\label{sec:engineering_evaluation}

Generated and reconstructed arrangements use the same class-based mass model for comparisons of $LCG$ and $KG$.
For functional region $i$, a represented mass is calculated from its occupied physical volume $V_i$ and the mass-intensity coefficient of its functional class $y_i$.
Fuel and ballast use representative physical liquid densities, while the remaining classes use fixed mass-intensity coefficients given in Table~\ref{tab:mass_proxies}.
Mass is placed at the physical centroid of each functional region before $LCG$ and $KG$ are calculated.

The non-liquid coefficients provide a common relative mass convention so that heavy machinery spaces contribute more strongly to the mass moments than cargo, stores, accommodation, or void.
For yachts, the cargo/mission class represents crew quarters and therefore uses the same relative value as accommodation.

The fixed-hull application extends this proxy model to whole-vessel equilibrium.
Fuel and ballast masses use their physical densities at a prescribed 98\% filling fraction.
The remaining functional masses are scaled by a dimensionless reference-specific factor $s_f$ while preserving their relative mass-intensity ratios:
\begin{equation}
m_i =
\begin{cases}
0.98\,\rho_{y_i}V_i,
& y_i \in \{\mathrm{fuel},\mathrm{ballast}\}, \\[3pt]
s_f\,\tilde{\rho}_{y_i}V_i,
& \text{otherwise},
\end{cases}
\end{equation}
where $\rho_{y_i}$ is the physical liquid density and $\tilde{\rho}_{y_i}$ is the proxy mass-intensity coefficient. 
The scale parameter $s_f$ adjusts the non-liquid functional masses relative to the reference displacement while preserving their relative values.

The total represented functional mass is
\begin{equation}
M_{\mathrm{comp}}=\sum_i m_i .
\end{equation}
The remaining vessel mass is represented by a permanent contribution such that
\begin{equation}
M_{\mathrm{comp}} + M_{\mathrm{perm}}
=
\Delta_{\mathrm{ref}},
\end{equation}
where $\Delta_{\mathrm{ref}}$ is the reference displacement.
The permanent contribution represents vessel mass not resolved by the functional-space model, including hull structure, outfit, and other unrepresented mass that cannot be determined from the functional arrangement alone.
The reference draught ratios are based on the design draught and depth of the available real-reference arrangements.
The functional-mass scale and the longitudinal and vertical centre of the permanent contribution were calibrated using a dedicated vessel subset spanning all six vessel families and were then held fixed during candidate evaluation.
The complete calibration procedure is reported in Appendix~\ref{app:gm_calibration}.

For each candidate, the combined functional and permanent mass moments provide the vessel $LCG$ and $KG$.
The hull hydrostatics are then solved for equilibrium draught and trim.
Longitudinal moment balance provides the forward and aft draughts and the physical solved trim defined in Section~\ref{sec:mass_stability_trim}.
At the solved condition, the hull hydrostatics provide $KB$ and $BM$, which combine with $KG$ to give $GM$.
The free-surface correction for the prescribed tank-loading condition is then applied to obtain $GM_{\mathrm{corr}}$.

\begin{figure}[t]
    \centering
    \includegraphics[width=0.8\linewidth]{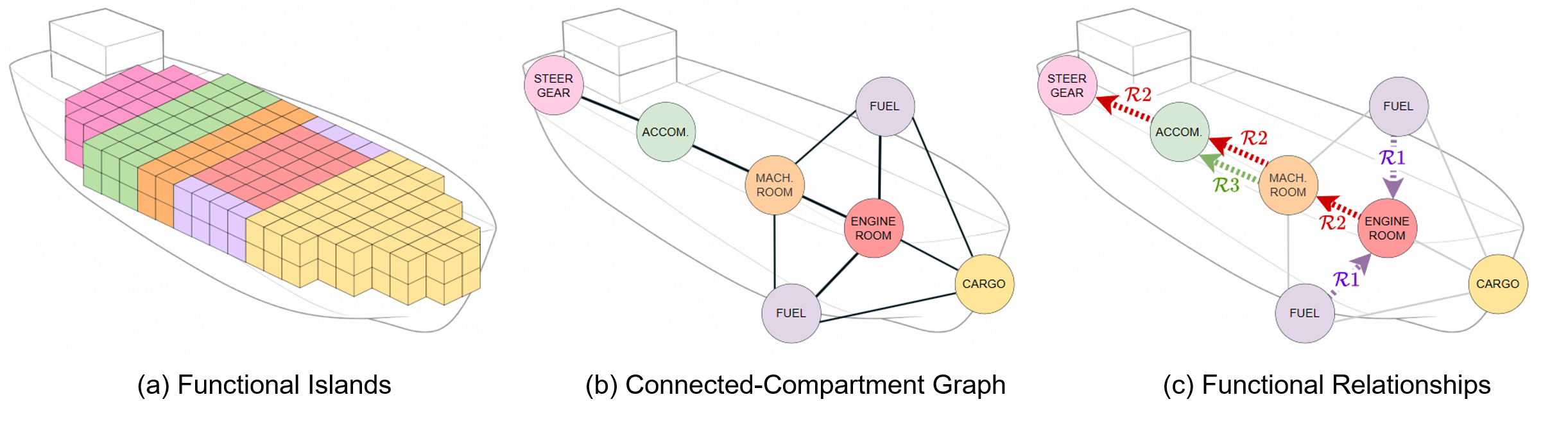}
    \caption{Construction of the CCGraph representation and the relationship set used in $J_{\mathrm{adj}}$.
    (a) A voxel arrangement is separated into connected regions with the same functional label, termed functional islands.
    (b) Each island becomes a node in the CCGraph, with edges representing shared physical interfaces.
    (c) The realised Fuel, Electrical, and Auxiliary relationships used in $J_{\mathrm{adj}}$ are shown between functional islands.
    Distances are evaluated as weighted shortest-path distances through the CCGraph.}
    \label{fig:jadj_construction}
    \vspace{-2.5pt}
\end{figure}

Functional adjacency cost ($J_{\mathrm{adj}}$) quantifies the topological separation between interdependent spaces at an architectural level.
It is evaluated in the same way for generated and reconstructed real arrangements and acts as a high-level proxy for functional connectivity efficiency.
The measure uses the CCGraph distance $d_G$ defined in Section~\ref{sec:functional_topology} to evaluate a prescribed set of functional relationships, $\mathcal{R}$.
These relationships are divided into three groups:
\begin{itemize}[topsep=1pt, itemsep=1pt]
\item \textit{$\mathcal{R}_1$--Fuel.} Connects each fuel space to its nearest engine-room space.
\item \textit{$\mathcal{R}_2$--Electrical.} Connects each machinery, accommodation, and steering-gear to its nearest engine-room space.
\item \textit{$\mathcal{R}_3$--Auxiliary.} Connects each accommodation space to its nearest machinery space.
\end{itemize}
Here, nearest is defined by the minimum CCGraph distance $d_G$.
Figure~\ref{fig:jadj_construction} illustrates the functional relationships and their representation on the CCGraph.

The functional adjacency cost is
\begin{equation}
J_{\mathrm{adj}}
=
\frac{1}{|\mathcal{R}|}
\sum_{(i,j)\in\mathcal{R}}
\frac{d_G(i,j)}{L}.
\end{equation}
A lower $J_{\mathrm{adj}}$ indicates less separation between the prescribed functional pairs.
Normalisation by $L$ provides a size-normalised measure for comparisons between vessels of different scale.
For the fixed-hull application, $L$ is constant and therefore does not affect the relative ordering of candidate arrangements.
In the fixed-hull application figures, the three relationship groups are shown as visual overlays to identify the functional pairs evaluated by $J_{\mathrm{adj}}$.

\subsection{Experimental design}
\label{sec:experimental_design}

The evaluation comprises three experiments addressing complementary properties required for design exploration: design-space plausibility, reference reconstruction, and requirement controllability.







\paragraph{Experiment 1: Design-space plausibility.}
Experiment~1 evaluates whether the generated design space aligns with the available real reference arrangements across the six vessel families.
This provides a family-level plausibility check on the arrangements produced by the generator.
The 50\,000 generated arrangements are compared with all 23 reconstructed references using functional-volume fractions, longitudinal and vertical functional-centroid locations, $LCG/L$, $KG/D$, and $J_{\mathrm{adj}}$.
A principal component analysis (PCA) provides a combined multivariate comparison using a 32-variable output descriptor.

The descriptor contains achieved $LCG/L$ and $KG/D$, zone-label frequencies for the nine active functional classes, mean normalised longitudinal and vertical locations ($c_x$, $c_z$) for each class, and normalised principal dimensions $L$, $B$, and $D$.
For classes absent from an arrangement, both spatial-location entries are set to zero.
Input targets, ship-type indicators, $J_{\mathrm{adj}}$, and $GM$ are excluded from the PCA descriptor.
The PCA basis is fitted to the generated arrangements, after which the reconstructed real references are projected into the same space.

\paragraph{Experiment 2: Reference reconstruction.}
Experiment~2 evaluates how closely the generator can reproduce one reconstructed real arrangement from each vessel family while retaining the fixed exterior reference envelope.
The experiment therefore tests the spatial expressivity of the representation and whether searching across different subdivisions provides closer matches to the reference than repeated assignment on one subdivision.
Reconstruction accuracy is defined as the percentage of voxels within the reference evaluation envelope that receive the same functional label as the reference:
\begin{equation}
A_{\mathrm{rec}}
=
100
\frac{
\sum_{v \in \mathcal{V}_{\mathrm{ref}}}
\mathbf{1}\!\left(\hat{y}_v = y_v\right)
}{
|\mathcal{V}_{\mathrm{ref}}|
},
\end{equation}
where $y_v$ and $\hat{y}_v$ denote the reference and reconstructed labels, respectively.
Void cells contained within the evaluation envelope are included.

Three ordinary-generation cases are compared.
Single draw is the mean accuracy across 100 independently generated candidates in which subdivision and assignment both vary.
Best fixed is the highest accuracy across 100 functional assignments on one generated spatial subdivision within the fixed reference envelope.
Best varying is the highest accuracy across 100 candidates in which both subdivision and functional assignment vary.
The fixed and varying searches therefore use separate candidate sets.


Mean absolute error (MAE) between two sets of corresponding quantities is defined as
\begin{equation}
\mathrm{MAE}=
\frac{1}{n}
\sum_{i=1}^{n}
\left|x_i-y_i\right|.
\label{eq:mae}
\end{equation}
Programme MAE applies this measure to the reconstructed and reference functional-volume fractions and is reported in percentage points.
Adjacency F1 compares the functional-class adjacency sets of the reconstructed and reference CCGraphs. Precision measures the fraction of reconstructed adjacencies present in the reference, recall measures the fraction of reference adjacencies recovered, and F1 is their harmonic mean.

A separate target-guided diagnostic tests how closely each sampled subdivision can reproduce the reference when the reference labels are used to guide relabelling. 
The procedure cycles through the subdivision zones and accepts a new label only when it is eligible, passes QC, and improves voxel-wise reconstruction accuracy. 
Because the subdivision geometry remains fixed, the diagnostic measures the additional agreement attainable within the sampled spatial structure.

\paragraph{Experiment 3: Requirement controllability.}
Experiment~3 is a one-factor sensitivity test of the generator's functional-volume controls.
It examines whether changing a requested functional volume produces a corresponding change in the generated arrangement.
This is important for design exploration because changes in design requirements should have a clear and traceable effect on the resulting layout.
For each vessel family, the seven requested functional-volume controls are varied independently within one fixed exterior hull envelope.
Each control is swept over four equally spaced levels spanning its family-specific range in Table~\ref{tab:sampling_ranges}, while the remaining six controls are held at their lower-bound values.
Void is not varied directly and remains the residual volume.

Each sweep is repeated across 100 independently sampled spatial subdivisions.
Within each subdivision, the assignment seed is held fixed across the four request levels so that the response to the changed requirement can be isolated.

The requested, effective, and achieved programme states defined in Section~\ref{sec:programme_assignment} are retained to identify where control is lost.
For each subdivision and control, ordinary least-squares slopes are fitted from requested to effective volume, $\beta_{\mathrm{eff}}$, and from requested to achieved volume, $\beta_{\mathrm{ach}}$.
A slope near one indicates an approximately proportional response, while a slope near zero indicates little response.
Achieved responses are classified as responsive ($0.50\leq\beta_{\mathrm{ach}}\leq1.15$), partial ($0.05\leq\beta_{\mathrm{ach}}<0.50$), over-response ($\beta_{\mathrm{ach}}>1.15$), or non-response ($\beta_{\mathrm{ach}}<0.05$).
Non-response is further separated according to whether the requested change is lost during capacity projection or during subsequent functional assignment.

\section{Results}
\label{sec:results}

The following sections report the main observations and outcomes from the three experiments.

\subsection{Design-space plausibility}
\label{sec:e1_population}

\begin{figure}[t]
\centering
\begin{subfigure}[t]{0.48\linewidth}
    \centering
    \includegraphics[width=\linewidth]{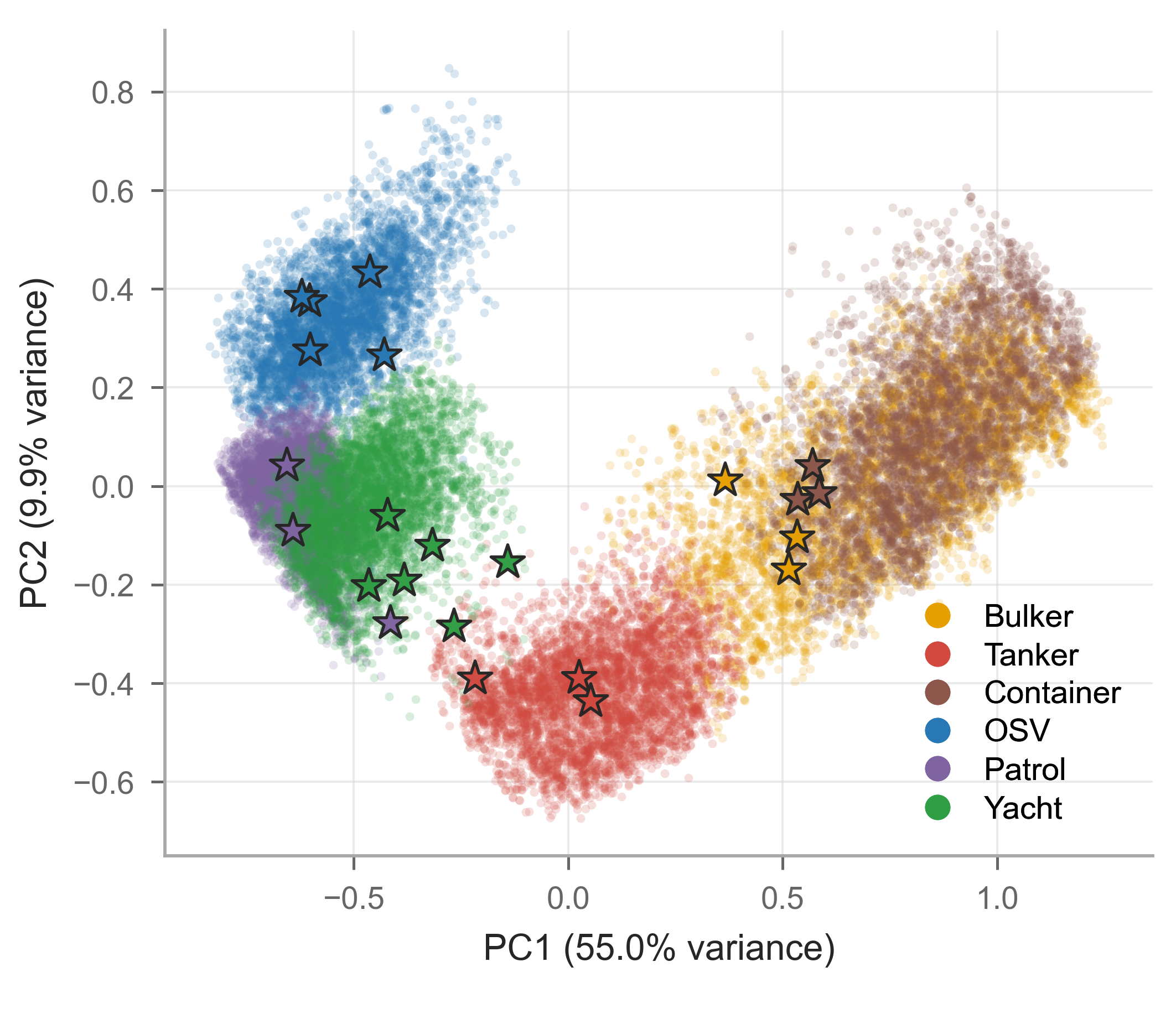}
    \caption{PCA projection of the generated arrangements and reconstructed real references.}
    \label{fig:e1_pca}
\end{subfigure}
\hfill
\begin{subfigure}[t]{0.48\linewidth}
    \centering
    \includegraphics[width=\linewidth]{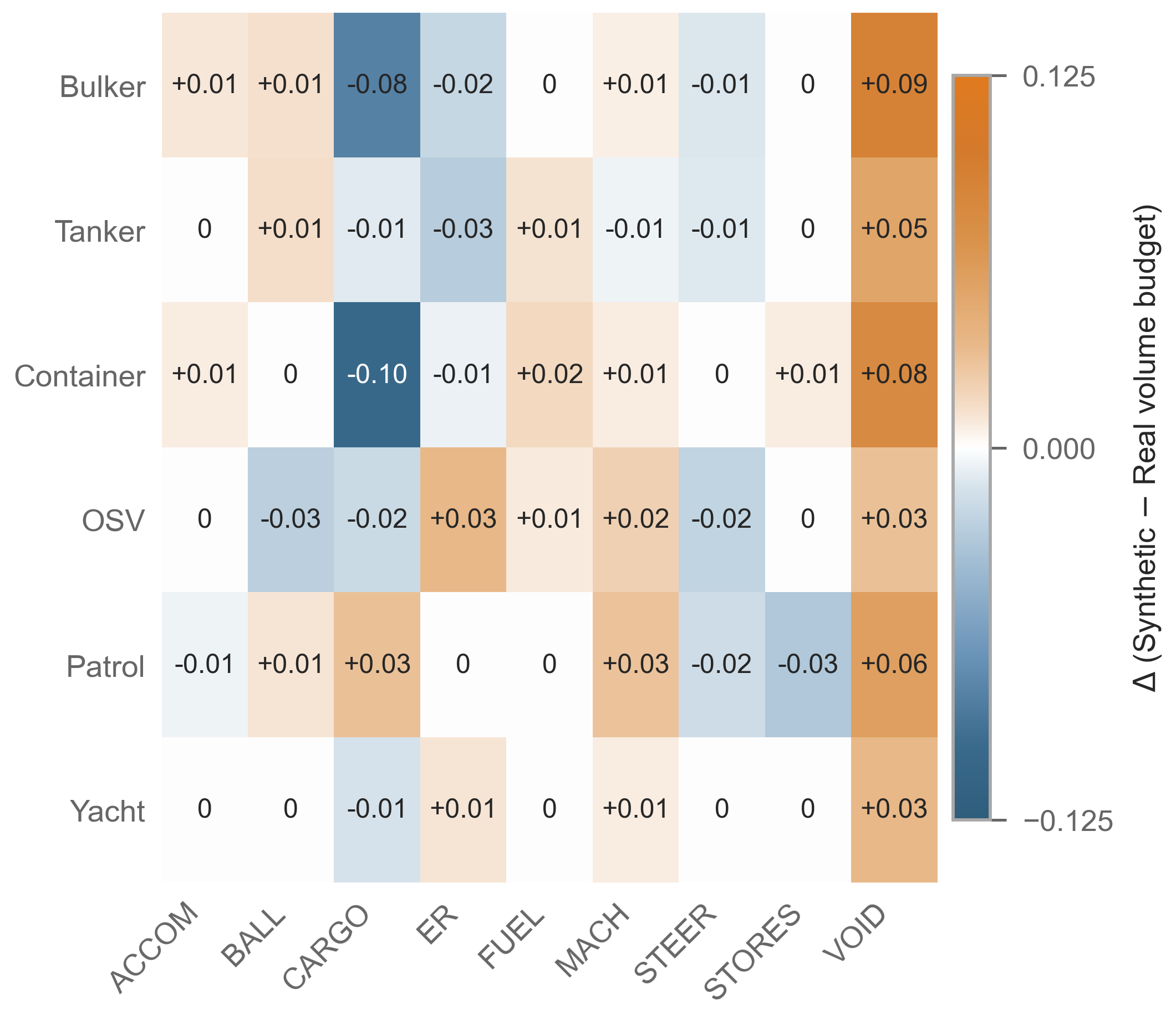}
    \caption{Difference between mean generated and real functional-volume fractions.}
    \label{fig:e1_budget}
\end{subfigure}

\vspace{4pt}

\begin{subfigure}[t]{0.48\linewidth}
    \centering
    \includegraphics[width=\linewidth]{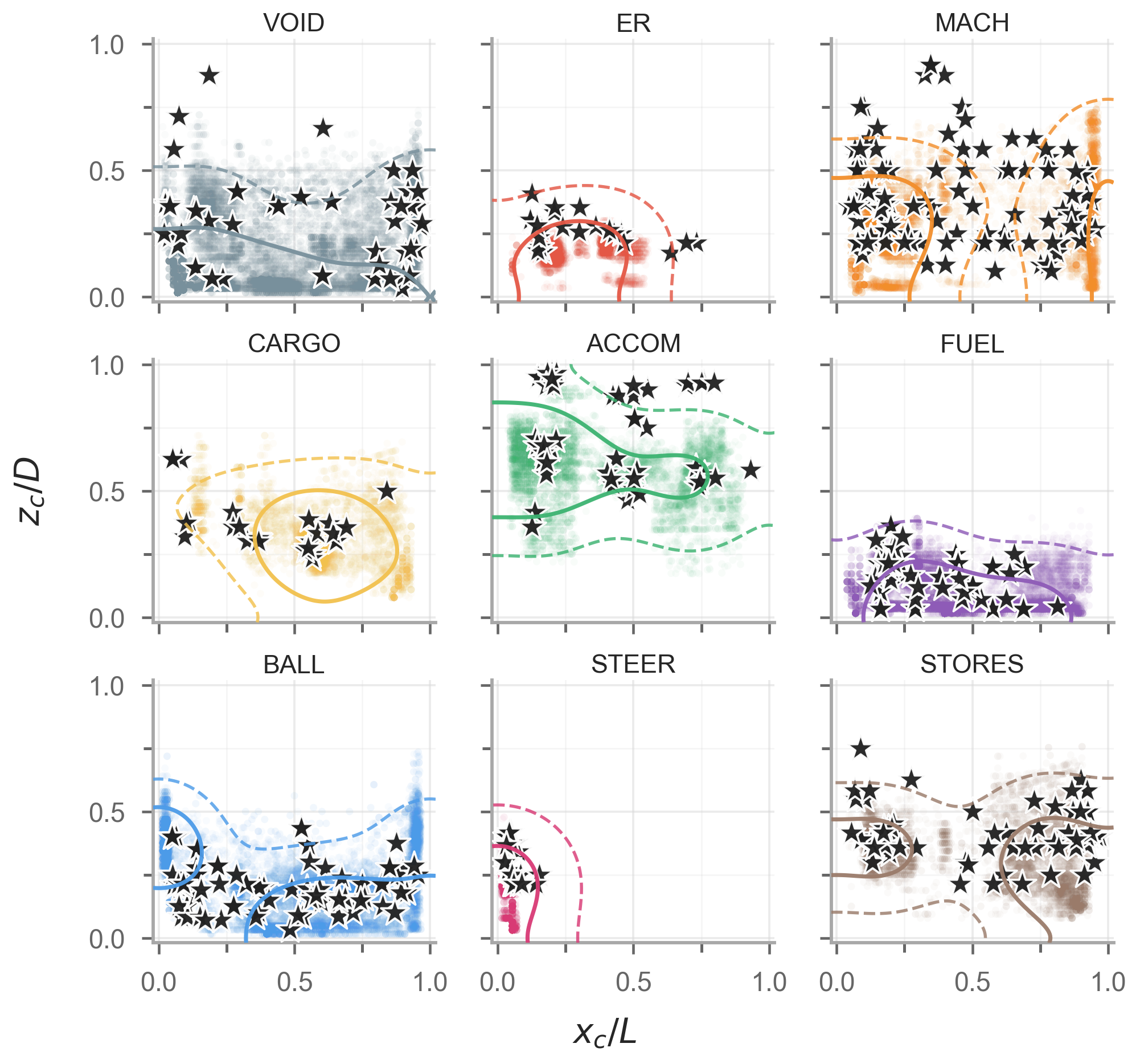}
    \caption{Longitudinal and vertical functional-centroid distributions.}
    \label{fig:e1_centroids}
\end{subfigure}
\hfill
\begin{subfigure}[t]{0.48\linewidth}
    \centering
    \includegraphics[width=\linewidth]{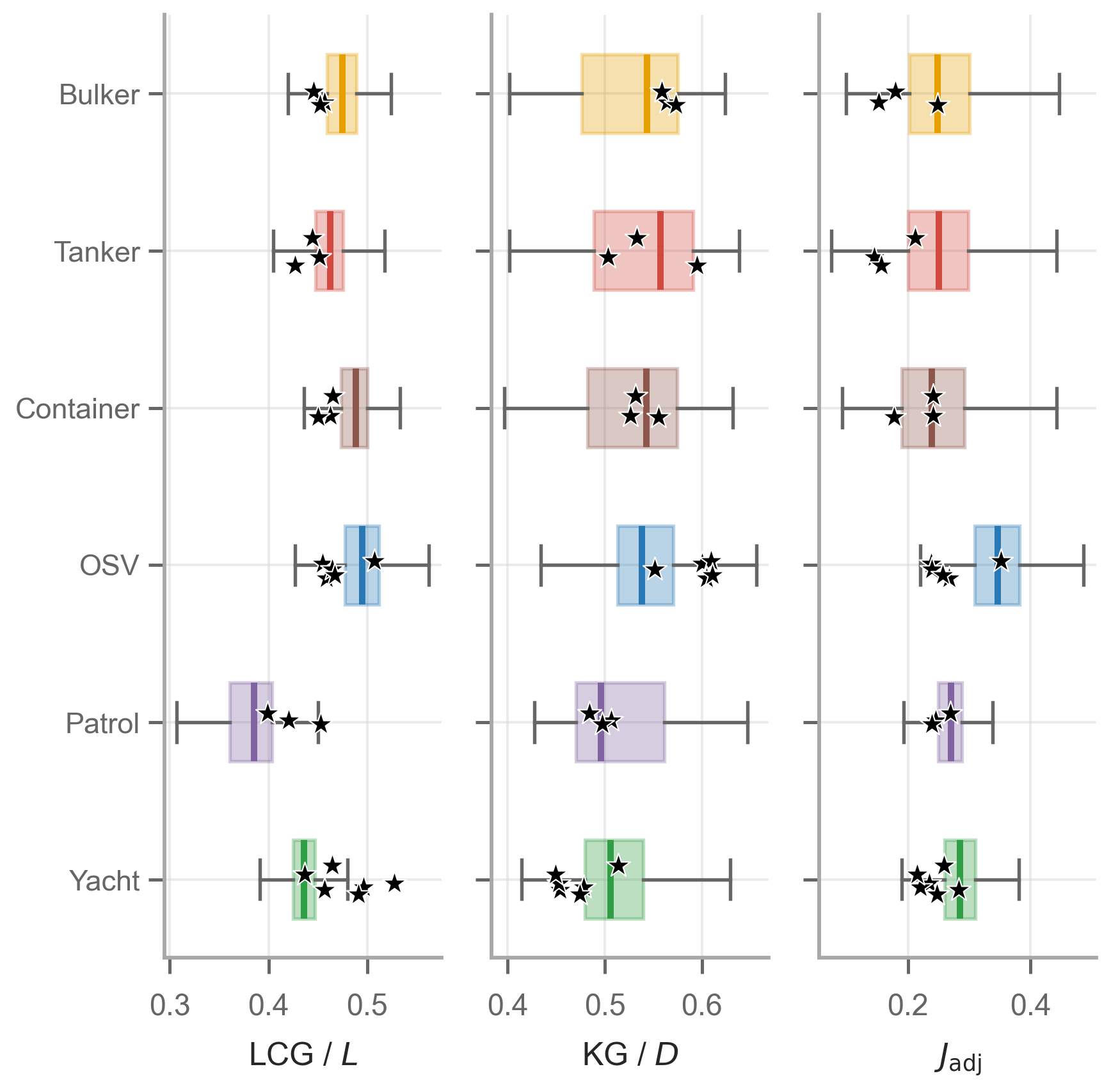}
    \caption{Family distributions of $LCG/L$, $KG/D$, and $J_{\mathrm{adj}}$, with real-reference values shown as stars.}
    \label{fig:e1_properties}
\end{subfigure}

\caption{Comparison of generated and reconstructed real arrangements across the six vessel families.
Where shown, circles ($\circ$) represent generated arrangements and stars ($\star$) indicate reconstructed real reference designs.}
\label{fig:e1_population_alignment}
\vspace{-10pt}
\end{figure}

The PCA projection in Figure~\ref{fig:e1_pca} shows distinct but overlapping regions for the six vessel families.
The first two principal components explain 53.7\% and 9.5\% of the total variance, respectively.
Visually, the real references generally lie within or near their corresponding generated regions, indicating good coverage of the available reference cases in the selected descriptor space.

Generated and real functional-volume fractions are close for most classes and families (Figure~\ref{fig:e1_budget}).
The clearest systematic difference is excess void in every vessel family, ranging from 0.03 to 0.09 of represented volume.
For bulk carriers and container vessels, this is accompanied by the largest cargo differences, with generated cargo fractions 0.08 and 0.10 below the respective real references.

The spatial comparison in Figure~\ref{fig:e1_centroids} reproduces the main functional placement patterns observed in the real arrangements, including aft engine and steering regions, low fuel and ballast, and higher accommodation.
The real references show a broader vertical range for some machinery and accommodation regions than the generated arrangements.

The real $LCG/L$ and $KG/D$ values also lie largely within or near the generated family distributions (Figure~\ref{fig:e1_properties}), although several yacht references extend beyond the generated $LCG/L$ range.
Real arrangements also tend toward lower $J_{\mathrm{adj}}$ than the generated distributions, most clearly for bulk carriers, tankers, and OSVs.

Overall, the comparison shows broad agreement between the generated design space and the available real reference cases while retaining identifiable family- and function-specific differences.

\subsection{Representation reconstruction}
\label{sec:e2_expressivity}

\begin{table}[t]
\centering
\scriptsize
\setlength{\tabcolsep}{3.6pt}
\renewcommand{\arraystretch}{1.2}
\begin{tabular}{lcccccc}
\hline
&
\multicolumn{4}{c}{\textbf{Reconstruction accuracy (\%)}} &
&
\\
\cline{2-5}
\textbf{Family} &
\textbf{Single draw} &
\textbf{Best fixed} &
\textbf{Best varying} &
\textbf{Target-guided$^\dagger$} &
\shortstack{\textbf{Programme}\\\textbf{MAE (pp)}} &
\shortstack{\textbf{Adjacency}\\\textbf{F1}} \\
\hline
Bulker
& 66.5 & 74.3 & \text{75.5} & 82.7 & 1.14 & 0.619 \\

Tanker
& 55.3 & 57.8 & \text{59.9} & 69.3 & 1.24 & 0.800 \\

Container vessel
& 61.4 & 66.5 & \text{69.1} & 77.1 & 1.07 & 0.783 \\

OSV
& 38.7 & 42.1 & \text{47.4} & 49.1 & 2.10 & 0.750 \\

Patrol vessel
& 18.2 & 21.1 & \text{23.0} & 36.7 & 4.47 & 0.611 \\

Yacht
& 41.3 & 46.0 & \text{50.6} & 58.3 & 1.58 & 0.783 \\
\hline
Average
& 46.9 & 51.3 & \text{54.3} & 62.2 & 1.93 & 0.724 \\
\hline
\end{tabular}

\vspace{2pt}
\parbox{0.98\linewidth}{\scriptsize
$^\dagger$Uses the real reference labels as a target-aware reconstruction diagnostic.
}
\vspace{2pt}
\caption{Fixed-envelope reconstruction of one real reference arrangement per vessel family.
Programme MAE and adjacency F1 refer to the accuracy-selected Best varying candidate.
The target-guided diagnostic uses the real reference labels and does not represent ordinary generator performance.}
\label{tab:e2_reconstruction}
\vspace{-10pt}
\end{table}

Table~\ref{tab:e2_reconstruction} shows that the best varying-subdivision reconstruction exceeds the single-draw baseline in all six cases.
Relative to this baseline, the best varying-subdivision reconstruction is higher by 4.6--9.3 percentage points.
Within the sampled searches, the best varying-subdivision candidate also exceeds the best fixed candidate in every case, with differences ranging from 1.2 percentage points for the bulker to 5.3 percentage points for the OSV.

However, reconstruction accuracy remains strongly case-dependent.
Best varying ranges from 23.0\% for the patrol vessel to 75.5\% for the bulker, with the tanker, container vessel, OSV, and yacht reaching 59.9\%, 69.1\%, 47.4\%, and 50.6\%, respectively.
Programme MAE is lowest for the container vessel at 1.07~pp and highest for the patrol vessel at 4.47~pp, while adjacency F1 ranges from 0.611 to 0.800.
Detailed reconstruction examples for the tanker and OSV are provided in Appendix Figure~\ref{fig:e2_examples}. 
These provide a visual comparison of the reference arrangement with the Best fixed, Best varying, and target-guided cases from multiple spatial and graph views.

Under the same 100-candidate search size, the varying-subdivision search finds a higher-accuracy candidate than the fixed-subdivision search for every vessel family, with gains of 1.2--5.3 percentage points.
Using the reference labels to guide relabelling increases accuracy by a further 1.7--13.7 percentage points across the six cases.
The resulting target-guided accuracies range from 36.7\% for the patrol vessel to 82.7\% for the bulker.
Because target guidance changes functional labels while retaining the sampled subdivision and assignment constraints, the remaining mismatch reflects the limits imposed by the sampled spatial representation.

\subsection{Requirement controllability}
\label{sec:e3_controllability}

\begin{figure}[t]
    \centering
    \begin{subfigure}[t]{0.32\linewidth}
        \centering
        \includegraphics[width=1.08\linewidth]{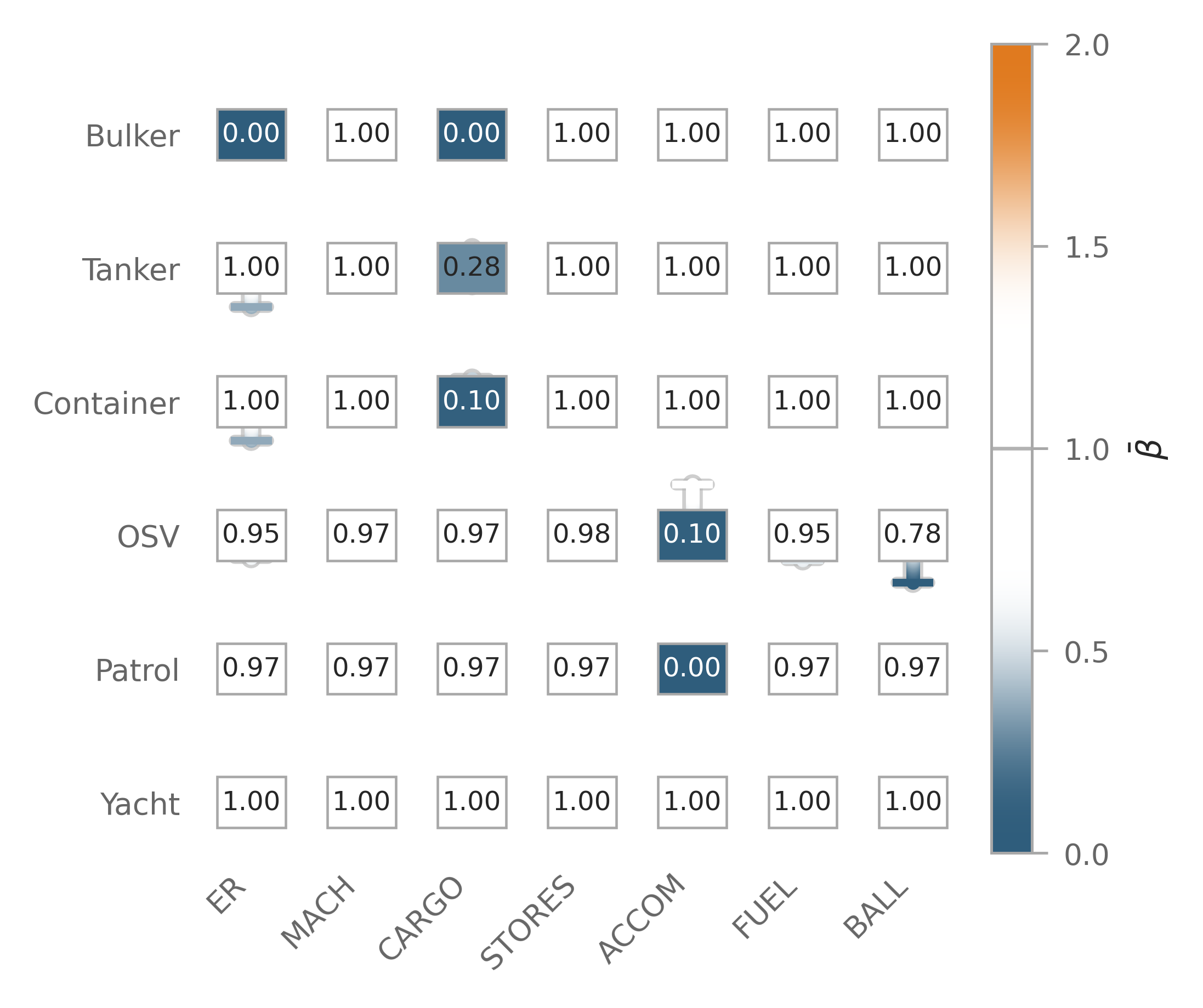}
        \caption{Median requested-to-effective response slope across 100 spatial subdivisions.}
        \label{fig:e3_effective_response}
    \end{subfigure}
    \hfill
    \begin{subfigure}[t]{0.32\linewidth}
        \centering
        \includegraphics[width=1.08\linewidth]{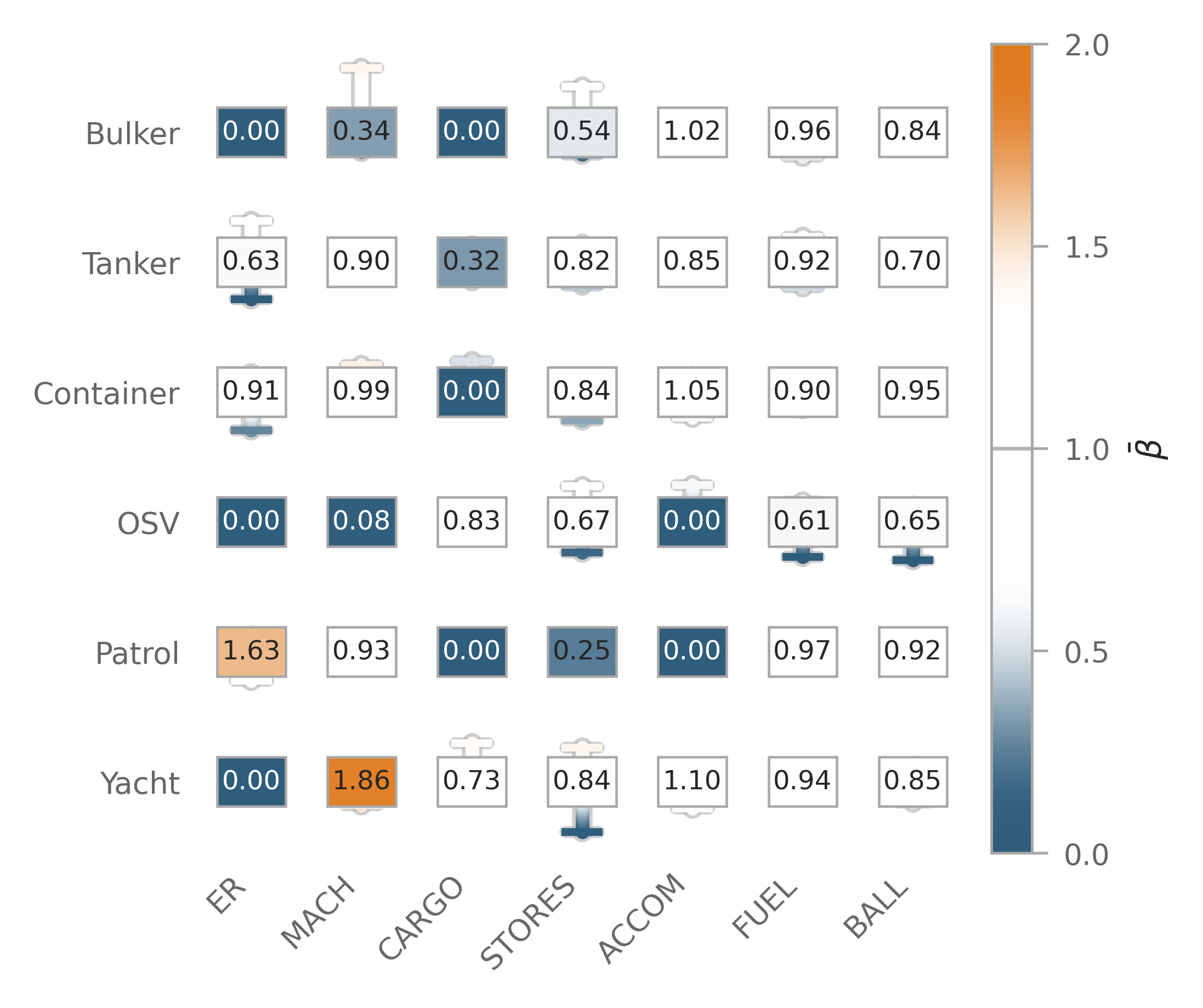}
        \caption{Median requested-to-achieved response slope across 100 spatial subdivisions.}
        \label{fig:e3_achieved_response}
    \end{subfigure}
    \hfill
    \begin{subfigure}[t]{0.32\linewidth}
        \centering
        \includegraphics[width=1.0\linewidth]{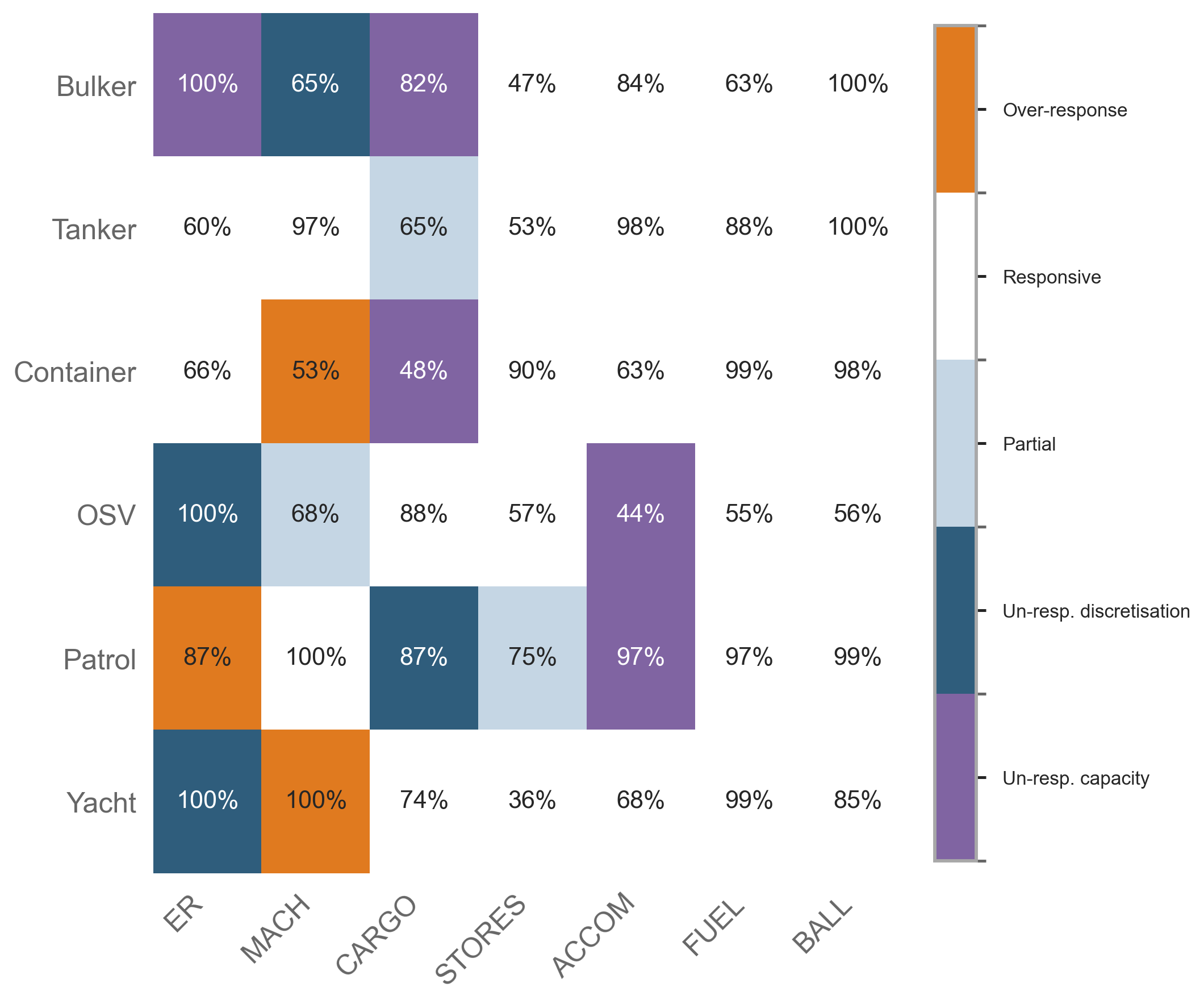}
        \caption{Dominant response class. Labels give the fraction of subdivisions in that class.}
        \label{fig:e3_response_classes}
    \end{subfigure}

    \caption{Requirement controllability under one-factor sensitivity sweeps across six vessel families and seven functional-volume controls.
    Panels~(a) and~(b) show the median requested-to-effective and requested-to-achieved response slopes.
    Panel~(c) shows the dominant response class across the 100 spatial subdivisions and separates non-response occurring during capacity projection from non-response occurring after projection.}
    \label{fig:e3_controllability}
\end{figure}

Figure~\ref{fig:e3_controllability} shows different response patterns across the functional controls.
Fuel and ballast retain the clearest response, with both median requested-to-effective and requested-to-achieved slopes remaining close to one for most vessel families.

Cargo in the transport families is mainly limited during capacity projection.
The median requested-to-effective cargo slopes are $\tilde{\beta}_{\mathrm{eff}}=0.00$ for the bulker, $0.28$ for the tanker, and $0.10$ for the container vessel.
The corresponding achieved slopes remain similarly low at $\tilde{\beta}_{\mathrm{ach}}=0.00$, $0.29$, and $0.09$.
The similar effective and achieved slopes place the main response loss at capacity projection under the realised subdivision and eligibility rules.

A different pattern occurs when the effective programme responds but the achieved programme remains low.
Bulker machinery changes from $\tilde{\beta}_{\mathrm{eff}}=1.00$ to $\tilde{\beta}_{\mathrm{ach}}=0.00$.
For the OSV, engine-room and machinery slopes fall from $0.95$ and $0.97$ to $0.00$ and $0.08$, respectively.
Patrol-vessel cargo and yacht engine-room volume show the same pattern, with effective slopes close to one and achieved slopes close to zero.
The drop from $\tilde{\beta}_{\mathrm{eff}}$ to $\tilde{\beta}_{\mathrm{ach}}$ places the main response loss during functional assignment, where the capacity-feasible programme is converted into discrete zone assignments.

Some controls respond by more than the requested change.
Container-vessel machinery increases from $\tilde{\beta}_{\mathrm{eff}}=1.00$ to $\tilde{\beta}_{\mathrm{ach}}=1.22$, patrol-vessel engine-room volume from $0.97$ to $1.63$, and yacht machinery from $1.00$ to $1.86$.
These cases are classified as over-response and result from discrete changes in the assigned zone volumes.

Panel~\ref{fig:e3_response_classes} summarises the dominant response class and its prevalence across the 100 subdivisions.
Capacity-limited non-response dominates bulker cargo, while non-response during functional assignment dominates OSV engine-room volume, patrol cargo, and yacht engine-room volume.
Over-response dominates patrol engine-room volume and yacht machinery.

\section{Application: fixed-hull concept exploration}
\label{sec:application}

The previous experiments characterise the generated design space and its limitations.
This section examines whether the framework can support early-stage comparison of alternative internal arrangements for one fixed vessel concept.
The application uses the same reconstructed OSV reference hull as Experiments~2 and~3.
An OSV was selected because several functions compete within a heterogeneous internal volume, making it a demanding arrangement case.
Its weaker reconstruction and requirement-response performance in Experiments~2 and~3 also make it a useful test of whether the generated alternatives still support engineering comparison.

The exterior hull, principal dimensions, and requested functional programme are held fixed while 500 alternative internal arrangements are generated.
The reference vessel has an overall length of 87.8~m, a moulded beam of 22.0~m, a moulded depth of 7.8~m, and a design draught of 5.5~m.
The requested volume fractions for engine room, machinery, cargo/mission space, stores, accommodation, fuel, and ballast are 9\%, 10\%, 20\%, 6\%, 21\%, 12\%, and 12\%, respectively.
These seven requested classes account for 90\% of the occupied hull volume, with the remaining capacity left unprescribed.
Steering gear is placed separately as a mandatory function, while void remains as residual capacity that provides slack during discrete functional assignment.

The resulting conceptual exploration is formulated as a constrained bi-objective selection over a generated candidate set:
\begin{equation}
\begin{aligned}
\min_{\mathbf{x}\in\mathcal{X}} \quad &
\left(
J_{\mathrm{adj}}(\mathbf{x}),
J_{\mathrm{trim}}(\mathbf{x})
\right) \\[1mm]
\text{subject to} \quad &
1 \leq GM_{\mathrm{corr}}(\mathbf{x}) \leq 2~\mathrm{m},\\
&
R_{\mathrm{prog}}(\mathbf{x}) \leq 8~\mathrm{pp},\\
&
\mathbf{x} \in \mathcal{X},
\end{aligned}
\label{eq:osv_application}
\end{equation}
where \(\mathbf{x}\) denotes a candidate internal arrangement and \(\mathcal{X}\) the sampled candidate set.
The first objective, \(J_{\mathrm{adj}}\), is the functional adjacency cost defined in Section~\ref{sec:engineering_evaluation}.
The second, \(J_{\mathrm{trim}}\), is the solved physical trim measure defined in Section~\ref{sec:mass_stability_trim}.
Together, they represent two arrangement-level considerations: functional spatial organisation and longitudinal mass balance.

Corrected metacentric height, \(GM_{\mathrm{corr}}\), is used as a constraint rather than an optimisation objective. Candidates are retained only when \(1~\mathrm{m} \leq GM_{\mathrm{corr}} \leq 2~\mathrm{m}\).
Functional-programme compliance uses the same absolute-error measure defined for MAE in Equation~\eqref{eq:mae}. Here, the errors are summed rather than averaged across the seven requested programme classes, with void excluded. This cumulative programme error is denoted \(R_{\mathrm{prog}}\), and candidates are restricted to \(R_{\mathrm{prog}}\leq8~\mathrm{pp}\).
Both bounds are application-specific selection criteria and are not proposed as universal requirements for ship internal arrangements.


\begin{figure}[t]
\centering
\includegraphics[width=0.6\linewidth]{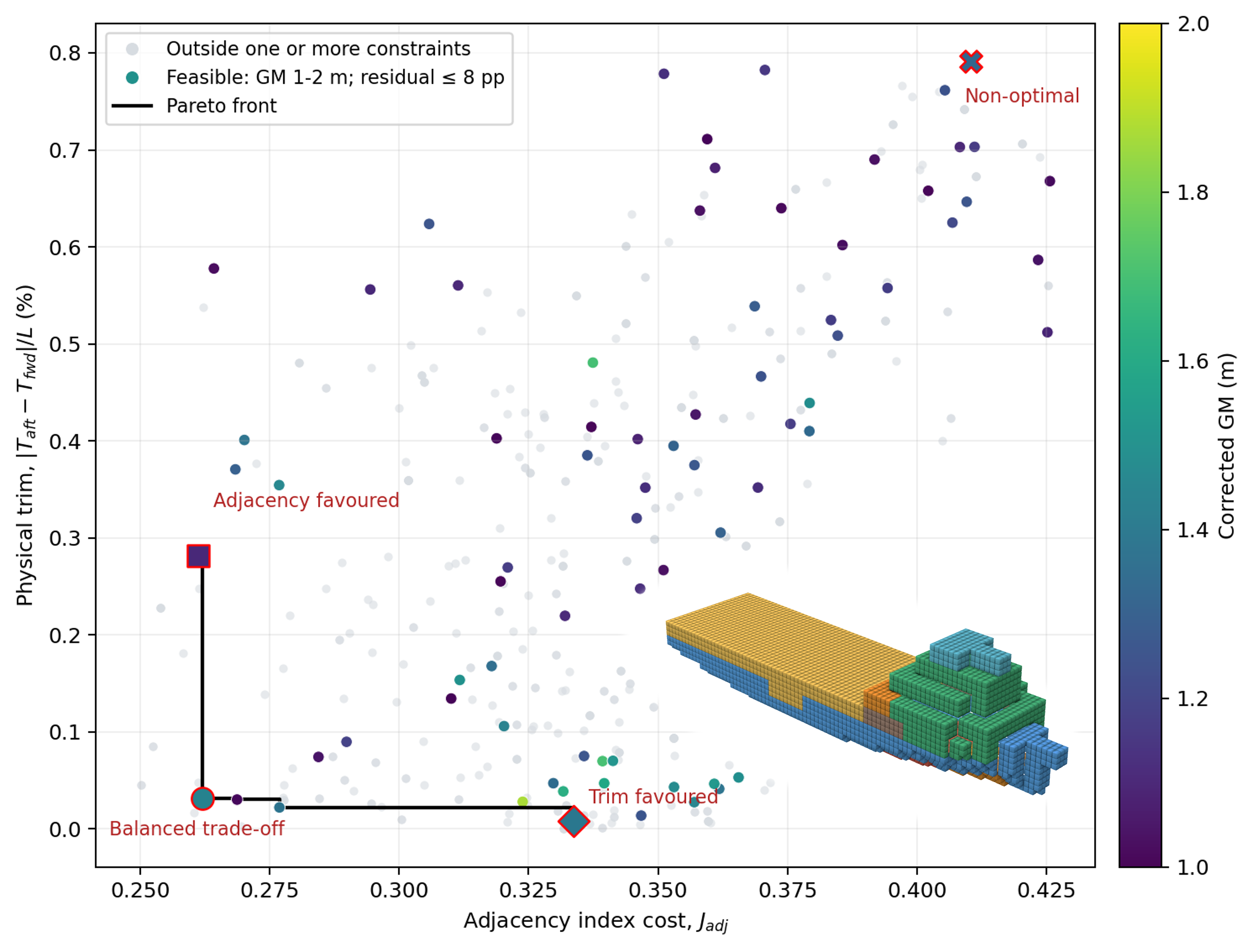}
\caption{Fixed-hull OSV concept exploration in the functional adjacency cost--trim design space. 
The same exterior envelope and design functional intent are retained across the generated alternatives. 
Grey candidates fall outside one or more application criteria. 
Feasible candidates are coloured by corrected GM, and the black line identifies the nondominated front.}
\label{fig:application_trade_space}
\vspace{-10pt}
\end{figure}

\subsection{Candidate results and trade-offs}
\label{sec:application_results}

All 500 generated arrangements produced usable representations and engineering evaluations.
Of these, 192 satisfy the selected corrected-\(GM\) band, and 119 also satisfy the programme-residual limit.
Five of the 119 feasible arrangements form the nondominated adjacency--trim set shown in Figure~\ref{fig:application_trade_space}.

The feasible concepts span 0.01--0.79\% of ship length in absolute trim.
Most candidates cluster toward relatively small trim values, while a smaller number show substantially larger longitudinal imbalance.
$J_{\mathrm{adj}}$ varies more continuously across the feasible population, producing a short trade-off front between lower functional adjacency cost and lower absolute trim.

\begin{figure}[t]
    \centering
    \begin{subfigure}[t]{0.49\linewidth}
        \centering
        \includegraphics[width=0.95\linewidth]{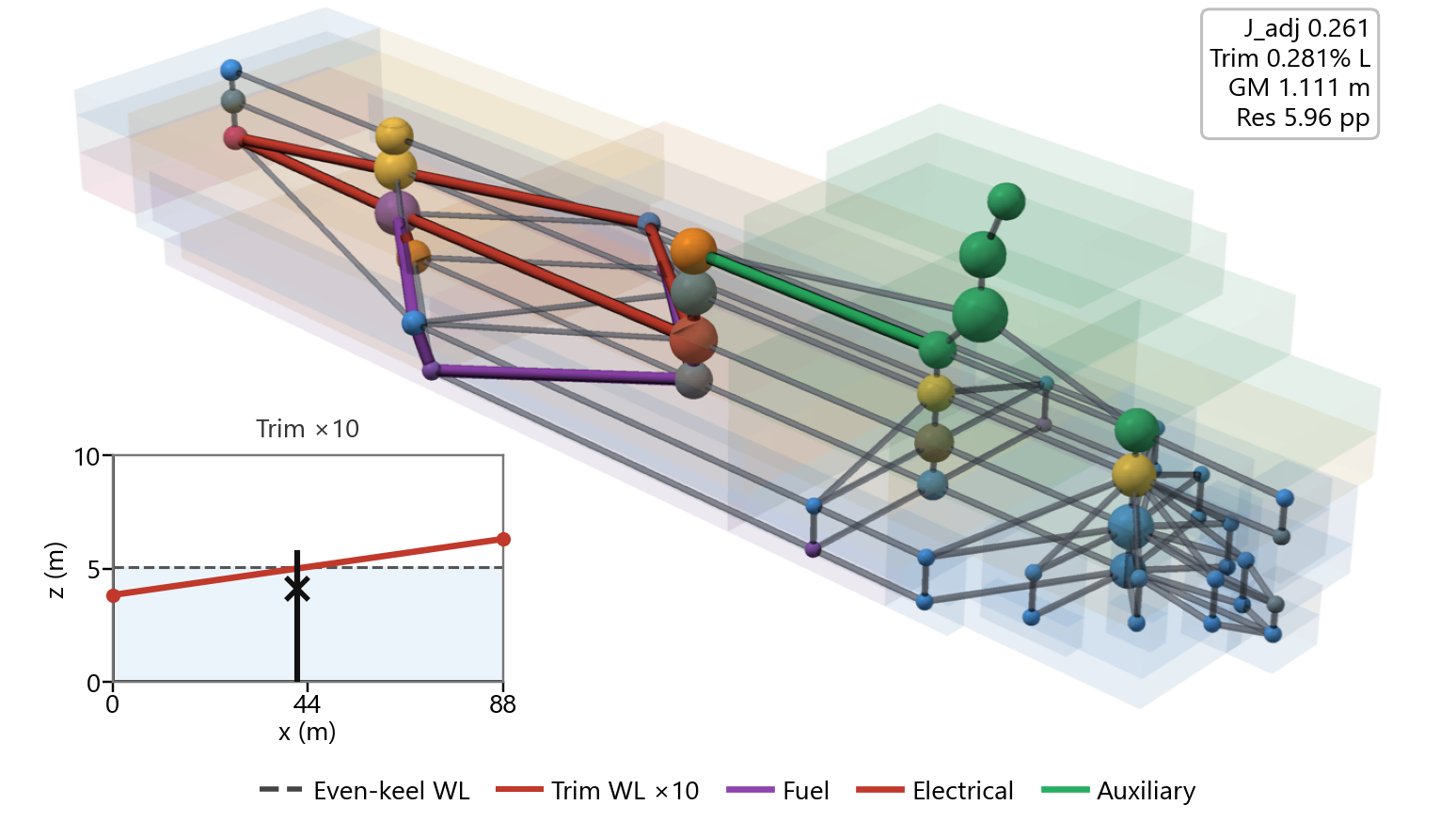}
        \caption{Adjacency-favoured concept.}
        \label{fig:application_concept_adjacency}
    \end{subfigure}
    \hfill
    \begin{subfigure}[t]{0.49\linewidth}
        \centering
        \includegraphics[width=0.95\linewidth]{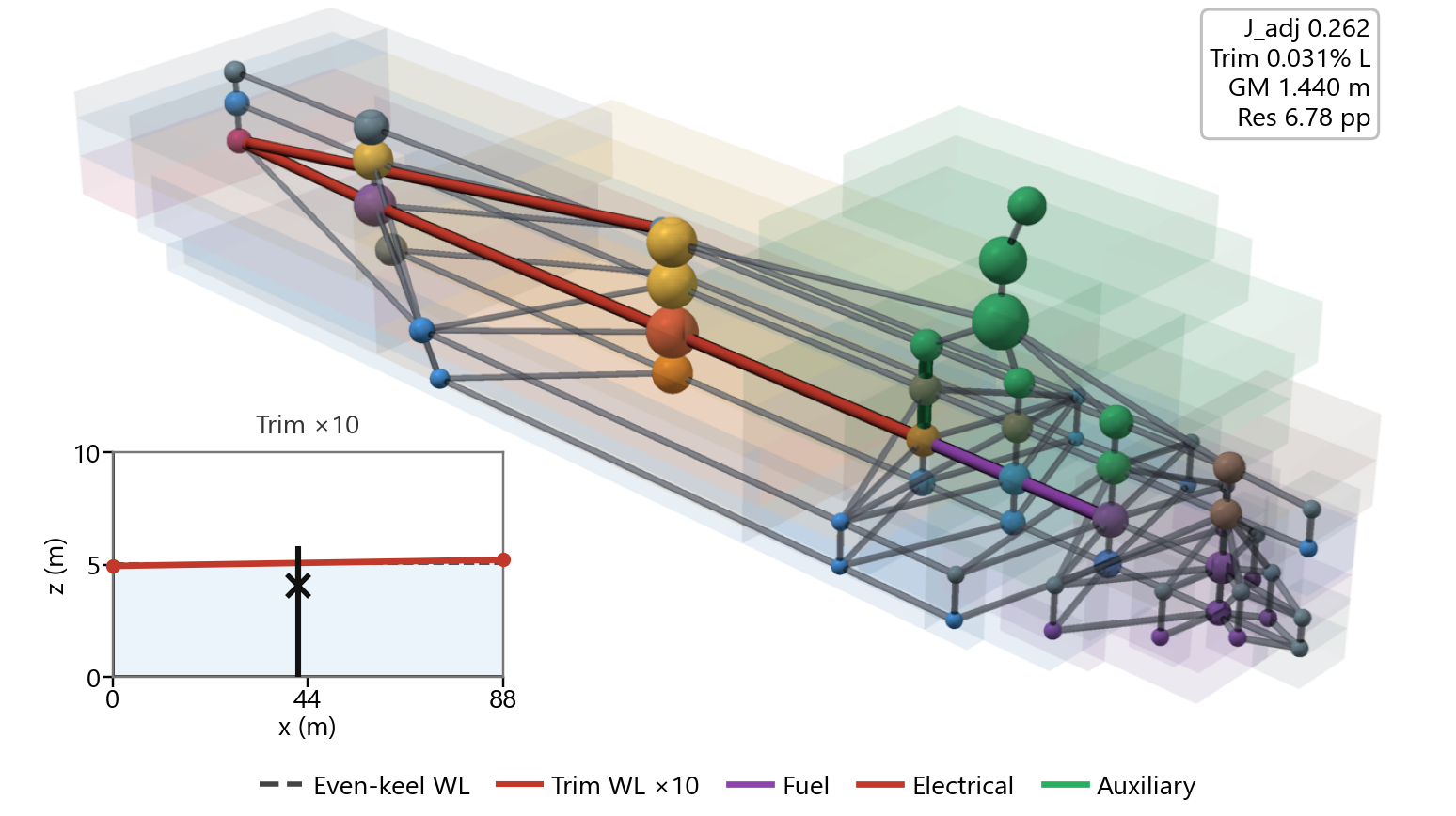}
        \caption{Balanced trade-off concept.}
        \label{fig:application_concept_balanced}
    \end{subfigure}
    \vspace{4pt}
    \begin{subfigure}[t]{0.49\linewidth}
        \centering
        \includegraphics[width=0.95\linewidth]{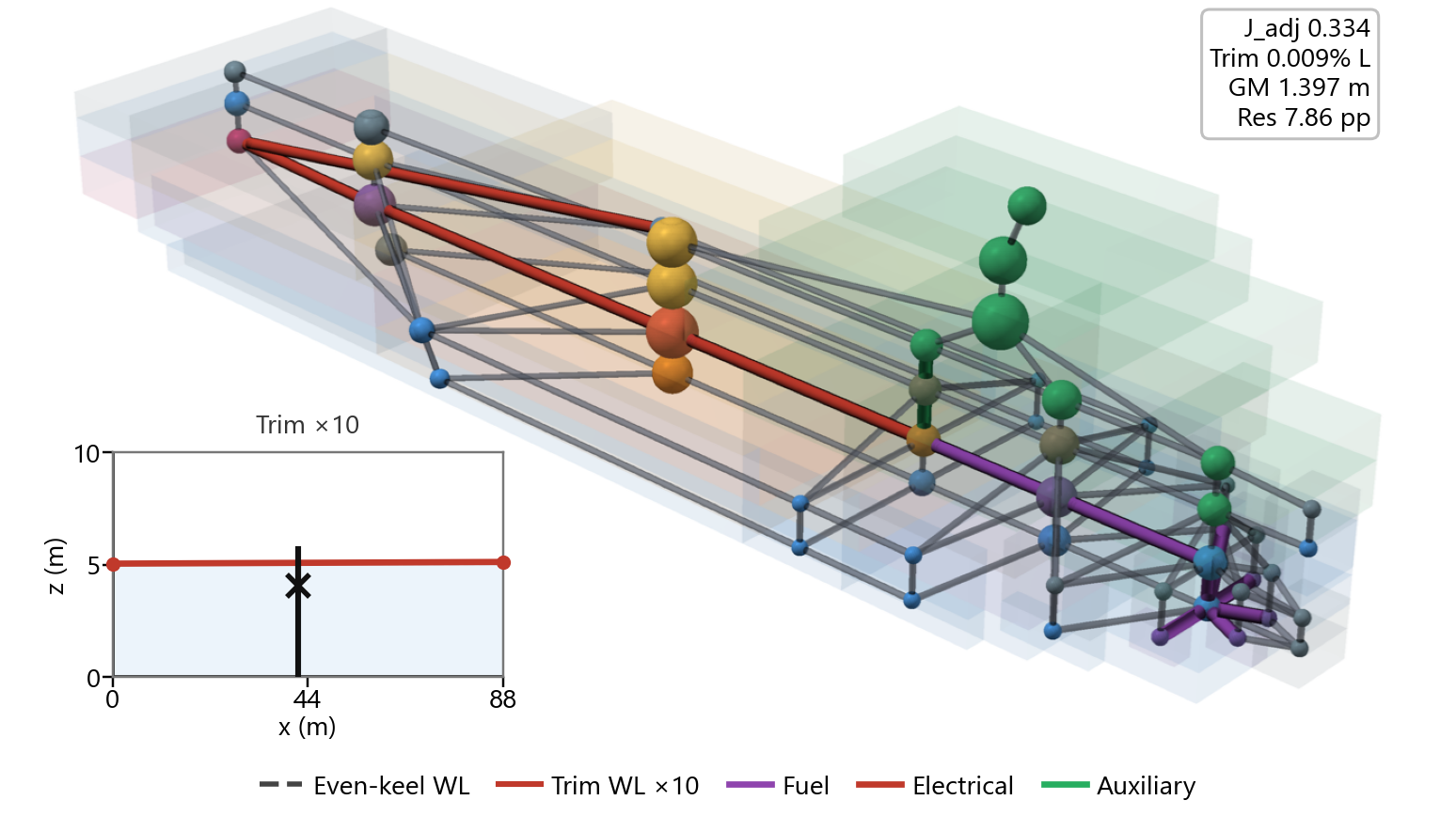}
        \caption{Trim-favoured concept.}
        \label{fig:application_concept_trim}
    \end{subfigure}
    \hfill
    \begin{subfigure}[t]{0.49\linewidth}
        \centering
        \includegraphics[width=0.95\linewidth]{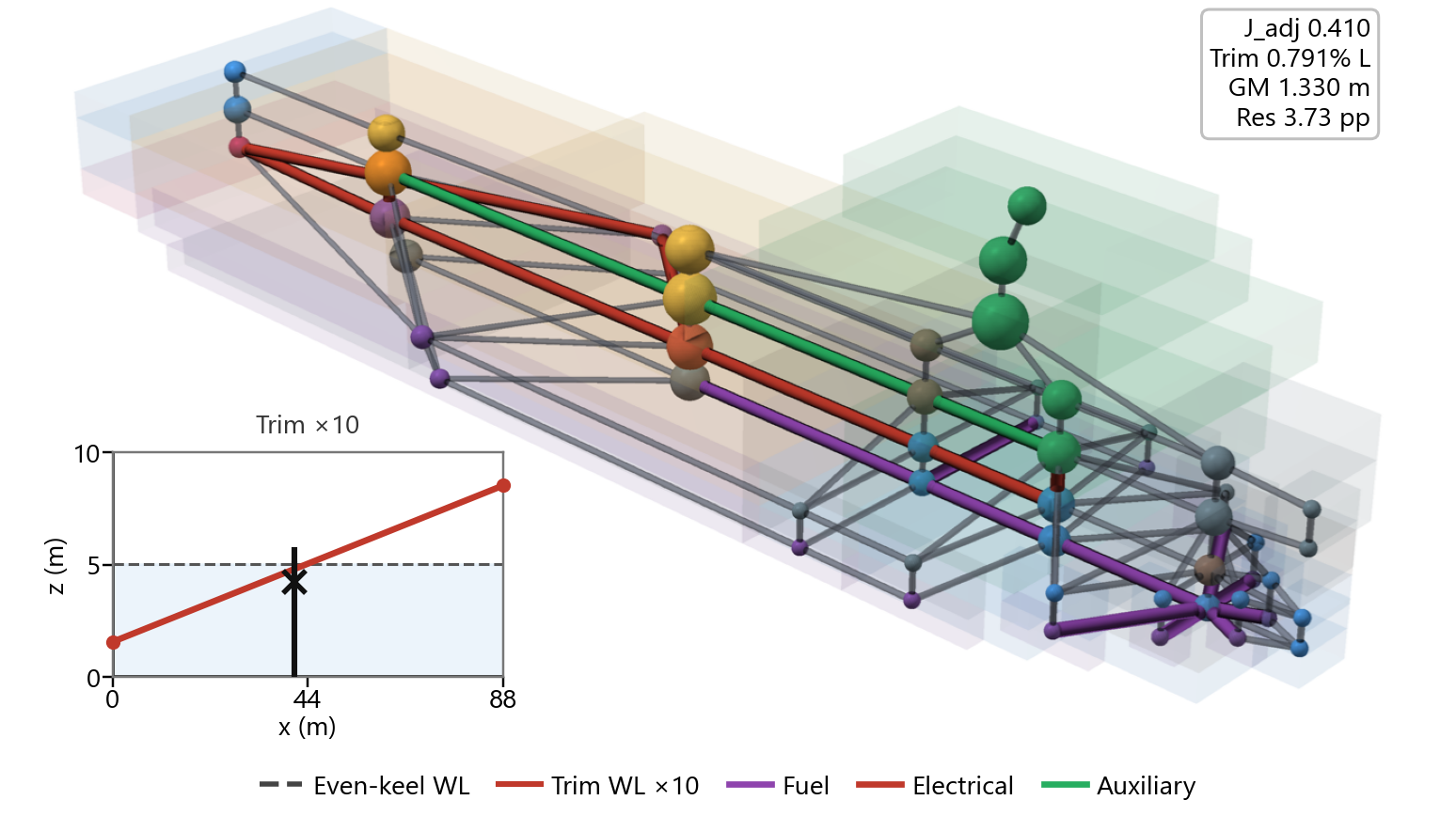}
        \caption{Feasible non-optimal comparison case.}
        \label{fig:application_concept_feasible}
    \end{subfigure}

    \caption{Selected arrangements from the fixed-hull OSV exploration:
    adjacency-favoured, balanced trade-off, trim-favoured, and a feasible
    non-optimal comparison case.
    The Fuel, Electrical, and Auxiliary overlays represent the functional
    relationship groups contributing to $J_{\mathrm{adj}}$.
    The trim inset shows the solved longitudinal waterline relative to the
    even-keel condition.}
    \label{fig:application_concepts}
    \vspace{-10pt}
\end{figure}

Figure~\ref{fig:application_concepts} shows four selected concepts from this trade space.
The adjacency-favoured arrangement reaches \(J_{\mathrm{adj}}=0.261\) with a trim magnitude of 0.281\% of ship length.
A nearby balanced concept retains almost the same adjacency cost, \(J_{\mathrm{adj}}=0.262\), while reducing trim to 0.031\%.
The trim-favoured arrangement reaches 0.009\% trim at a higher adjacency cost of 0.334.
By comparison, the feasible non-optimal example performs poorly in both objectives, with \(J_{\mathrm{adj}}=0.410\) and 0.791\% trim.

The selected arrangements differ in their spatial organisation.
The adjacency-favoured concept gives shorter graph separation for the prescribed functional relationships, while the trim-favoured concept shows a more even longitudinal distribution of functional volume.
The feasible non-optimal case combines higher functional separation with poorer longitudinal balance.
Within one fixed vessel envelope and requested functional programme, the generated alternatives therefore produce distinct adjacency--trim outcomes and a small nondominated set.
This shows that the framework can support early-stage comparison between arrangement concepts before a detailed layout is selected.

\section{Discussion}
\label{sec:discussion}

The experiments and application reveal four different properties of the generated design space, including whether it resembles the available real arrangements, what arrangements it can represent, how well its inputs control the final layout, and whether the resulting spatial variation matters for engineering design.
The results show that these properties should be considered separately.

Sample set agreement can hide restrictions in what the representation can express.
The limited improvement obtained from target-guided reconstruction in some cases suggests that choosing better functional labels cannot overcome all of the mismatch with a real arrangement.
The subdivision, eligibility rules, and discrete zone structure already determine part of what can be represented.
The comparison of programme and adjacency measures points to a related distinction where reproducing the amount and approximate location of each function appears easier than reproducing how those functions are organised relative to one another.
Relational organisation is therefore another property of the arrangement rather than a consequence of matching its functional volumes.

The requirement sweeps show that control is also filtered through the generation procedure.
A requested functional volume first depends on whether the realised subdivision provides enough eligible capacity and then on whether the assignment procedure can realise that capacity using sufficient discrete zones.
The requested functional fractions are therefore best interpreted as requests to the generator rather than independent continuous coordinates of the final arrangement.
Requested and achieved programmes should consequently remain separate when the generated data are interpreted or used for later learning tasks.

The cargo results suggest one consequence of this restricted capacity.
The families with the largest cargo deficits relative to the real references also show weak cargo response during capacity projection.
This supports the hypothesis that some geometric volume exists within the hull but is unavailable to cargo under the current subdivision and eligibility rules.
The same mechanism may contribute to residual void when available space cannot be used by the function requiring additional capacity.
The present experiments do not determine how much of the excess void is caused by this mechanism, so it remains a convincing factor rather than a complete explanation.

The fixed-hull application shows why the spatial variation that remains is still important.
Meeting the same hull and functional requirements does not determine a single engineering outcome.
How functions are placed and connected changes both their functional relationships and the longitudinal mass distribution.
Internal arrangement is therefore a design variable in its own right, and generating multiple spatial alternatives exposes engineering choices that functional-volume compliance alone cannot show.

Together, the generated design space is partly shaped by the procedure used to create it.
The subdivision and eligibility rules define which spatial opportunities are available, while the assignment procedure determines how those opportunities are used.
These restrictions explain why some arrangements and requested changes are easier to realise than others.
At the same time, the remaining variation is large enough to produce different engineering outcomes.
The experiments therefore show both the useful design space produced by the framework and the mechanisms that currently limit it.

Beyond these procedural restrictions, several broader limitations remain:
\begin{itemize}[topsep=2pt, itemsep=1pt]

\item \textit{Limited data.} Only 23 reconstructed real arrangements are available, and they are not a statistical sample of the wider fleet. 
They also informed the manually defined generator ranges and spatial preferences, rather than being used to derive those rules statistically.

\item \textit{Simplified representation.} Both real and generated arrangements are reduced to common functional classes and graph/voxel representations. 
This removes finer room-level geometry, access routes, distributed systems, irregular boundaries, etc.

\item \textit{Dataset dependence.} The 50,000 arrangements all come from the same generator and QC process, so the dataset inherits its sampling choices and modelling assumptions. 
This matters particularly for later machine-learning use.

\item \textit{Conceptual engineering evaluation.} The compartment mass properties and unresolved vessel mass are approximate and calibrated at family level. 
They should be further tuned/validated against detailed real vessel data where available.
The study also does not include damage stability, detailed regulatory compliance, structure, systems routing, access, or production constraints. 
The OSV application only identifies the best trade-offs within the 500 sampled alternatives, not a globally optimal arrangement.
\end{itemize}

\section{Conclusions}
\label{sec:conclusions}

This study developed and evaluated a common procedural framework for generating 3D internal arrangements across six vessel families.
The resulting 50\,000 arrangements align with the available real references, but the experiments also show clear limits in what the current subdivision and assignment procedure can represent and control.
The fixed-hull OSV application shows that arrangements with the same hull and requested functional programme can produce different engineering outcomes.
Spatial arrangement therefore matters beyond simply meeting volume requirements.

Procedural generation can provide the large number of alternatives needed for systematic design-space analysis, but the resulting data inherit the assumptions and restrictions of the procedure that creates them.
This study shows that these effects can be examined by testing not only plausibility, but also what layouts can be represented, how requested changes affect the final arrangement, and whether the resulting spatial differences affect engineering outcomes.

Future work will focus on more flexible subdivision, spatial information learned from real general arrangements, and assignment methods that consider the arrangement more globally.
A larger set of real arrangements will also be needed for broader validation.
The reproducible generation process and dataset are released publicly to support later learning-based studies~\cite{odendaal2026shipspacedataset,odendaal2026shipspacega}.



\bibliographystyle{unsrt}  
\bibliography{references}

\newpage
\appendix

\section{Compartment assignment rules}
\label{app:compartment_assignment}

Functional assignment separates hard placement constraints from soft spatial preferences.
First eligibility is assigned and then mandatory compartments are placed.
For the remaining eligible zones, the spatial multiplier $P^{\mathrm{space}}_{iq}$ in Section~\ref{sec:programme_assignment} modifies the greedy assignment score according to zone location and type.
A value of $P^{\mathrm{space}}=1$ is neutral, values above unity favour a placement, and values below penalise it.
The preference multipliers are bounded so that the spatial prior remains a soft influence on the greedy score rather than a hard placement rule.
The upper bound of $2.0$ limits the strength of a preferred location, while the non-zero lower bounds retain feasible but spatially less-preferred placements.
The numerical preference functions for each functional class and vessel family are given in Table~\ref{tab:app_spatial_preferences}.
Coordinates are normalised as $x=x_i/L$ and $z=z_i/D$, with $x=0$ aft, $x=1$ forward, and $z=0$ at the keel.

{\scriptsize
\begin{longtable}{
L{0.15\textwidth}
L{0.19\textwidth}
L{0.58\textwidth}}
\toprule
\textbf{Function} &
\textbf{Family} &
\textbf{Spatial preference $P^{\mathrm{space}}$} \\
\midrule
\endfirsthead

\toprule
\textbf{Function} &
\textbf{Family} &
\textbf{Spatial preference $P^{\mathrm{space}}$} \\
\midrule
\endhead

Engine room &
Bulker, Tanker, Container &
$\mathrm{clip}[1.5(1-x),\,0.1,\,2.0]$,
evaluated only within the engine-region eligibility
($x/L < 0.30$). \\

&
OSV, Patrol, Yacht &
$\mathrm{clip}[0.5+1.2x,\,0.1,\,2.0]$,
evaluated only within engine-room-eligible zones. \\[2pt]

Machinery &
Bulker, Tanker, Container, OSV, Patrol &
$\mathrm{clip}\!\left[
0.6+1.4\max(0,1-|x-0.20|/0.30),
\,0.2,\,2.0\right]$. \\

&
Yacht &
$\mathrm{clip}\!\left[
0.6+1.4\max(0,1-|x-0.40|/0.30),
\,0.2,\,2.0\right]$. \\[2pt]

Cargo / mission &
Bulker, Tanker, Container &
$1+0.2(1-z)$,
multiplied by $1.6$ in lower-main zones and $1.3$ in double-bottom zones;
final value clipped to $[0.8,2.0]$. \\

&
OSV, Patrol &
$\displaystyle
\mathrm{clip}\!\left[
\frac{[1+0.6(1-x)][1+0.4z]}{1.3},
\,0.8,\,2.0\right]$. \\

&
Yacht &
$\mathrm{clip}[1+0.2(1-z),\,0.8,\,2.0]$. \\[2pt]

Stores &
All &
$\displaystyle
\mathrm{clip}\!\left[
\frac{(0.4+1.2a)(0.8+0.4z)}{1.1},
\,0.1,\,2.0\right]$,
where
$a=\max[0,1-\max(0,x-0.40)/0.40]$. \\[2pt]

Accommodation &
Bulker, Tanker, Container &
$0.4+z$;
in engine-upper zones,
$\max(0.4+z,\,1.25+0.6z)$;
final value clipped to $[0.3,2.0]$. \\

&
OSV, Patrol &
$2.0$ in main-upper zones;
$1.5$ in superstructure zones;
otherwise $0.5+0.5z$. \\

&
Yacht &
$2.2$ in main-upper zones (clipped to $2.0$);
$2.0$ in engine-upper zones;
in superstructure deck $d_{\mathrm{SS}}=0,1,\ldots$,
$\mathrm{clip}[1.55-0.35d_{\mathrm{SS}},\,0.3,\,2.0]$;
otherwise $0.5+0.5z$, clipped to $[0.3,2.0]$.
Here $d_{\mathrm{SS}}=0$ denotes the lowest superstructure deck. \\[2pt]

Fuel tanks &
All &
For $x\leq0.55$,
$x_p=1+g(1-x/0.55)$,
with $g=0.75$ in side zones and $0.50$ otherwise.
For $x>0.55$,
$x_p=\max[g_f,1-(x-0.55)/0.45]$,
with $g_f=0.15$ in side zones and $0.20$ otherwise.
The vertical term is
$z_p=1+0.35(1-z)$ in lower side zones and
$z_p=1+0.10(1-z)$ otherwise.
Then
$P^{\mathrm{space}}=\mathrm{clip}(x_pz_p/1.1,\,0.1,\,2.0)$. \\[2pt]

Ballast tanks &
All &
$1.8$ in double-bottom zones;
$1.3$ in lower zones;
otherwise $\max[0.3,1-0.4z]$.
Outside peak zones this is multiplied by $1+0.2(1-x)$.
In the fore peak it is multiplied by $\max[0.35,1-0.55z]$.
Final value clipped to $[0.1,2.0]$. \\[2pt]

Steering gear &
All &
Not scored; placement is fixed by the mandatory aft-peak rule. \\

Void &
All &
Not scored during greedy functional assignment. \\

\bottomrule
\caption{Numerical spatial-preference multipliers used during greedy
functional assignment.
The functions are evaluated only after zone eligibility and mandatory
pre-placement have been applied.}
\label{tab:app_spatial_preferences}
\vspace{-10pt}
\end{longtable}
}

\begin{figure}[H]
\centering
\includegraphics[width=0.96\linewidth]{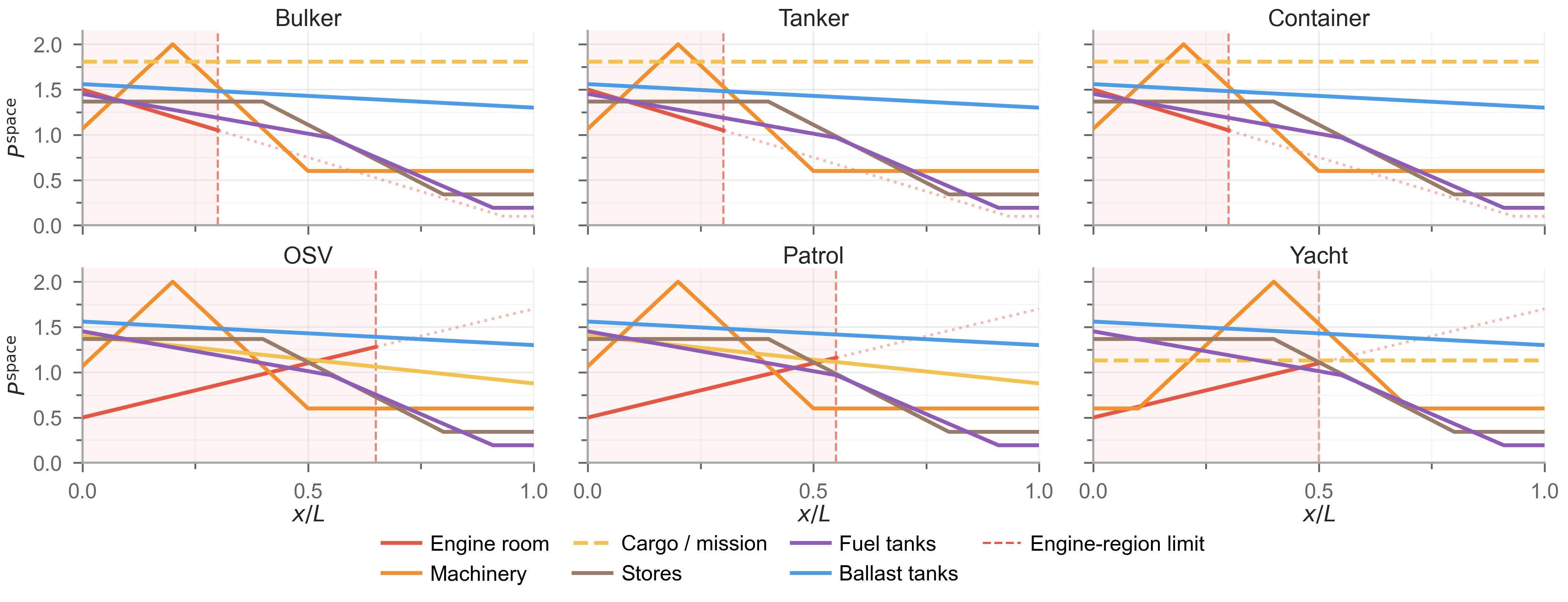}
\caption{Numerical spatial-preference multipliers used during greedy
functional assignment for each ship type.}
\label{fig:spatial_preference}
\vspace{-5pt}
\end{figure}

The spatial preference functions are family-level heuristic priors.
Their directional trends were informed by inspection of the available real general arrangements, while the numerical multipliers were iteratively tuned during generator development to reduce recurrent implausible placements.
The encoded preference priors for each vessel family are visualised in Figure~\ref{fig:spatial_preference}.
\vspace{-5pt}

\section{Production hull library}
\label{app:stl_provenance}

The production hull library contains 17 processed, cleaned and converted hull templates used to define the physical envelopes of generated arrangements.
Sixteen templates are derived from geometries obtained from the \href{https://www.cgtrader.com/3d-models/watercraft/industrial-watercraft/hulls-collection-2010a682-0584-4725-9da9-e8740bdf8883}{\emph{Hulls Collection}} distributed through CGTrader,
and one uses the \href{http://www.simman2008.dk/KCS/kcs_link.htm}{KRISO Container Ship (KCS) benchmark hull}.
Table~\ref{tab:app_stl_provenance} summarises the provenance and use of the production hull templates.

Before use, the source geometries are converted to STL where required, aligned to the common coordinate convention, symmetrised where necessary, and scaled to the sampled principal dimensions before voxelisation.
The original source geometry objects are not redistributed with the released dataset.
Only the derived arrangement representations and associated metadata are provided.

{\scriptsize
\begin{longtable}{
L{0.10\textwidth}
L{0.20\textwidth}
L{0.30\textwidth}
L{0.30\textwidth}}
\toprule
\textbf{Hull ID} &
\textbf{Family use} &
\textbf{Source / provenance} &
\textbf{Licence / reuse status} \\
\midrule
\endfirsthead

\toprule
\textbf{Hull ID} &
\textbf{Family use} &
\textbf{Source / provenance} &
\textbf{Licence / reuse status} \\
\midrule
\endhead

Transport 1--4 &
Bulker / tanker / container vessel &
\href{https://www.cgtrader.com/3d-models/watercraft/industrial-watercraft/hulls-collection-2010a682-0584-4725-9da9-e8740bdf8883}
{CGTrader, \emph{Hulls Collection}} &
Royalty Free licence; source geometries not redistributed \\

Transport 5 &
Bulker / tanker / container vessel &
\href{http://www.simman2008.dk/KCS/kcs_link.htm}
{KRISO Container Ship (KCS) benchmark hull} &
Publicly available benchmark geometry; source geometry not redistributed \\

Work 1--6 &
OSV &
\href{https://www.cgtrader.com/3d-models/watercraft/industrial-watercraft/hulls-collection-2010a682-0584-4725-9da9-e8740bdf8883}
{CGTrader, \emph{Hulls Collection}} &
Royalty Free licence; source geometries not redistributed \\

Yacht 1--3 &
Yacht &
\href{https://www.cgtrader.com/3d-models/watercraft/industrial-watercraft/hulls-collection-2010a682-0584-4725-9da9-e8740bdf8883}
{CGTrader, \emph{Hulls Collection}} &
Royalty Free licence; source geometries not redistributed \\

Patrol 1--3 &
Patrol vessel &
\href{https://www.cgtrader.com/3d-models/watercraft/industrial-watercraft/hulls-collection-2010a682-0584-4725-9da9-e8740bdf8883}
{CGTrader, \emph{Hulls Collection}} &
Royalty Free licence; source geometries not redistributed \\

\bottomrule
\caption{Provenance and use of the 17 production hull templates.
All geometries are aligned to the common coordinate convention,
symmetrised where necessary, scaled to the sampled principal dimensions,
and voxelised before use.
Original source hull geometries are not redistributed with the released dataset.}
\label{tab:app_stl_provenance}
\end{longtable}
}

\section{Reconstructed real-reference corpus}
\label{app:real_ga_sources}

\begin{figure}[H]
\centering
\includegraphics[width=0.97\linewidth]{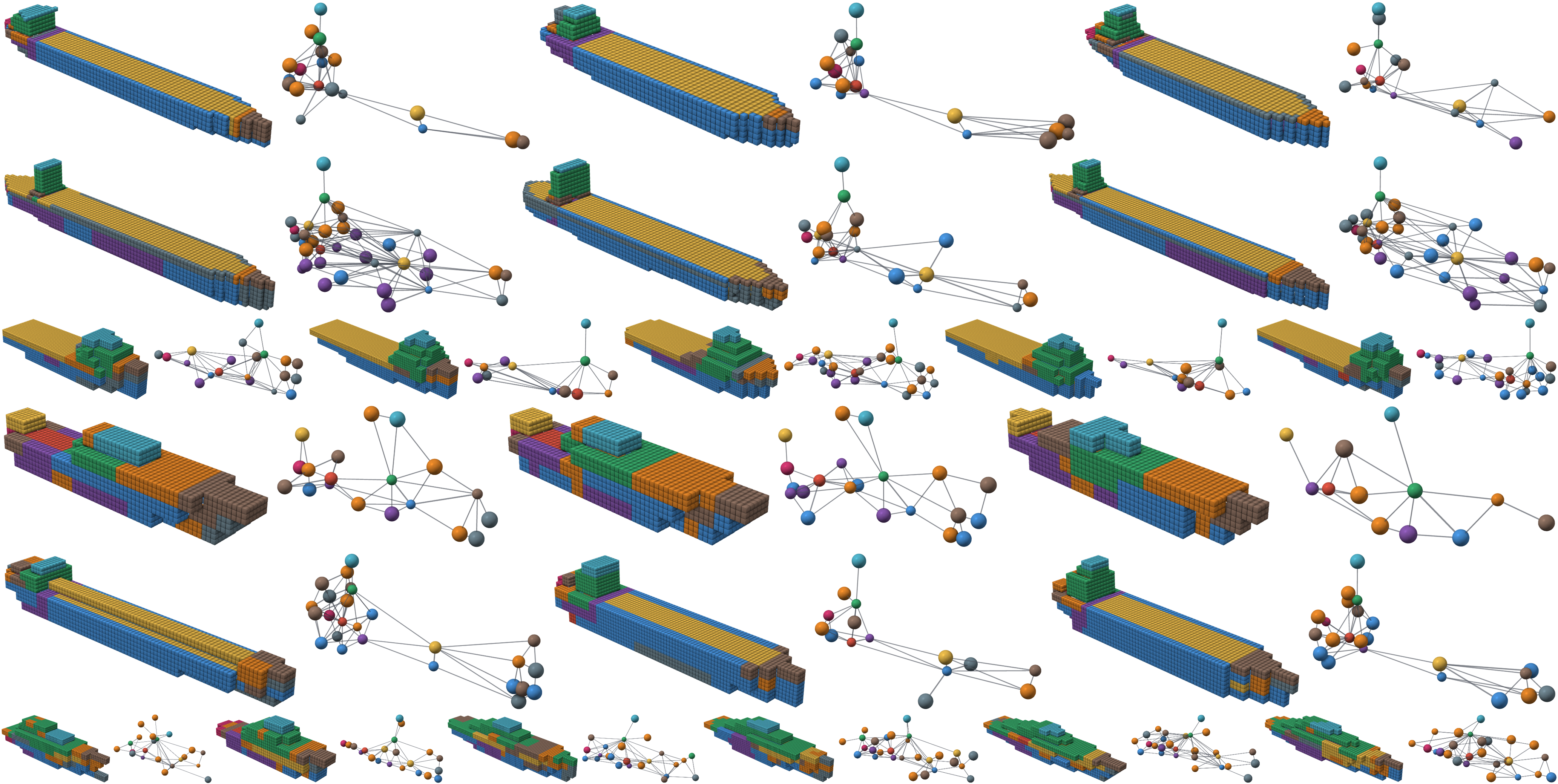}
\vspace{3pt}
\caption{Reconstructed real-reference corpus used in the study.
Each vessel is represented using the labelled three-dimensional encoding and
derived CCGraph used for comparison with generated arrangements.}
\label{fig:sup_reference_collection}
\vspace{-5pt}
\end{figure}

The real-reference corpus contains 23 manually reconstructed general arrangements covering the six vessel families considered in this study.
It comprises six yachts, five offshore support vessels, and three each of bulk carriers, tankers, container vessels, and patrol vessels.
These arrangements provide real-design reference cases for population comparison and the fixed-envelope experiments described in the main paper.
Figure~\ref{fig:sup_reference_collection} shows the complete reconstructed corpus using the common representations adopted in this study.

The source material was compiled from publicly available vessel documentation, published ship-reference volumes, academic theses, and online vessel databases.
The drawings vary in scale, annotation, geometric detail, and compartment terminology.
Each reference was therefore manually reconstructed as a labelled three-dimensional arrangement and mapped to the common functional taxonomy used by the generator.
The reconstructed models retain the spatial information required for comparison but should not be interpreted as exact CAD reproductions of the source vessels.

Table~\ref{tab:app_ga_sources} records the provenance and experimental use of each reconstructed reference.
All 23 arrangements are used as plausibility anchors in Experiment 1~(\ref{sec:e1_population}).
One reference from each vessel family is additionally used in the fixed-envelope reconstruction Experiment 2~(\ref{sec:e2_expressivity}) for direct comparison.
Experiment 3~(\ref{sec:e3_controllability}) uses a fixed real-reference envelope for each family. 
For the OSV, the same reconstructed reference and hull envelope are used in Experiment 2, Experiment 3, and the fixed-hull engineering application~(\ref{sec:application}).

For reproducibility purposes, the release contains the derived three-dimensional representations, the graph representations, and the associated metadata.

{\scriptsize
\begin{longtable}{
p{0.10\textwidth}
p{0.13\textwidth}
p{0.70\textwidth}}

\toprule
\textbf{ID} &
\textbf{Family} &
\textbf{Source / provenance} \\
\midrule
\endfirsthead

\toprule
\textbf{ID} &
\textbf{Family} &
\textbf{Source / provenance} \\
\midrule
\endhead
Y01 &
Yacht &
\href{https://www.burgessyachts.com/en/buy-a-yacht/yachts-for-sale/boardwalk-v-00008893}
{Burgess, \emph{Boardwalk} (Feadship), general arrangement and deck plans}. \\


Y02 &
Yacht &
\href{https://www.yachtcharterfleet.com/luxury-charter-yacht-48099/deep-blue-layout.htm}
{YachtCharterFleet, \emph{Deep Blue} (Amels), general arrangement and deck plans}. \\

Y03 &
Yacht &
\href{https://www.yachtcharterfleet.com/luxury-charter-yacht-49216/o-ptasia-layout.htm}
{YachtCharterFleet, \emph{O'Ptasia} (Golden Yachts), general arrangement and deck plans}. \\

Y04 &
Yacht &
\href{https://www.yachtcharterfleet.com/luxury-charter-yacht-27048/quattroelle-layout.htm}
{YachtCharterFleet, \emph{Quattroelle} (Lürssen), general arrangement and deck plans}. \\

Y05 &
Yacht &
\href{https://www.yachtcharterfleet.com/luxury-charter-yacht-53600/seven-seas-layout.htm}
{YachtCharterFleet, \emph{Seven Seas} (Oceanco), general arrangement and deck plans}. \\

Y06 &
Yacht &
\href{https://www.yachtcharterfleet.com/luxury-charter-yacht-49014/sophia-layout.htm}
{YachtCharterFleet, \emph{Sophia} (Feadship), general arrangement and deck plans}. \\

\midrule

O01 &
OSV &
\href{https://camgsa.mx/assets/files/BA_Centurion_Brochure.pdf}
{CAMGSA, \emph{Centurión}, supply vessel, vessel brochure and general arrangement}. \\

O02 &
OSV &
\href{https://www.mmaoffshore.com/theme/mmaoffshorecomau/assets/public/File/vessel-specs/MMA_Leeuwin.pdf}
{MMA Offshore, \emph{MMA Leeuwin}, platform supply vessel, vessel specification and general arrangement}. \\

O03 &
OSV &
\href{https://www.mmaoffshore.com/theme/mmaoffshorecomau/assets/public/File/vessel-specs/MMA_PRIDE.pdf}
{MMA Offshore, \emph{MMA Pride}, vessel specification and general arrangement;
Cybermarine Technologies drawing CT-18-5524-0002, Rev.~01 (2018)}. \\

O04 &
OSV &
\href{https://www.neptunems.com/vessel-fleet/mma-pinnacle}
{MMA Offshore, \emph{MMA Pinnacle}, vessel documentation and general arrangement}. \\

O05 &
OSV &
\href{https://www.vard.com/shipbuilding/references/skandi-responder}
{VARD, \emph{Skandi Responder}, vessel reference and general arrangement;
drawing 808-101-001, yard no.~808}. \\

\midrule

B01 &
Bulker &
\href{https://onebulk.net/wp-content/uploads/2019/09/ZZ-Vessel-Presentation-NORDIC-LONDON.pdf}
{T.K. Shipbuilding Co., Ltd., \emph{Nordic London}, 34,000~DWT bulk carrier,
vessel presentation and general arrangement}. \\

B02 &
Bulker &
\href{https://www.orient.nl/wp-content/uploads/2021/11/General-Arrangement-Nanjing-Confidence.pdf}
{Nanjing Dongze Shipyard Co., Ltd., \emph{Nanjing Confidence},
general arrangement}. \\

B03 &
Bulker &
\href{https://oelsm.com/wp-content/uploads/2025/06/P-04-General-Arrangement-LUNAR-STAR-1.pdf}{Oshima Shipbuilding Co., Ltd., 37,200/46,700~DWT open bulk carrier,
general arrangement; drawing no.~10040000, 31 October 1995}. \\

\midrule

T01 &
Tanker &
\href{https://springmarines.com/uploads/media/Document/2022/07/general-arrangement.pdf}
{Besiktas Shipyard, 7,000~DWT product tanker, general arrangement;
Armada Voyager Six Co.~Ltd., drawing RST22M110.1}. \\

T02 &
Tanker &
\href{https://springmarines.com/en/fleet-detail/zumrut-ana}
{Spring Marine, \emph{Zumrut Ana}, oil/chemical tanker,
vessel documentation and general arrangement}. \\

T03 &
Tanker &
\href{https://www.nedamaritime.gr/img/media_room/brochures/NedaAframaxNewBuildings.pdf}
{Neda Maritime, \emph{Seriana}/\emph{Stellata}, Aframax tankers,
newbuilding documentation and general arrangement}. \\

\midrule

C01 &
Container &
\href{https://www.naviecapitani.it/Navi%20e%20Capitani/gallerie%20navi/Containers/foto/N/Northern%20Power/Piano%20Generale.pdf}
{Daewoo Shipbuilding \& Marine Engineering, NRS 4,600~TEU containership,
general arrangement; hull no.~4140, drawing 101Z029}. \\

C02 &
Container &
\href{https://vinnen.com/fleet/}
{F.~A.~Vinnen \& Co., \emph{Leonidio}, 4,957~TEU containership,
fleet documentation and general arrangement}. \\

C03 &
Container &
\href{https://www.naviecapitani.it/gallerie%20navi/containers/foto/M/MSC/MSC%20Florida/Piano.pdf}
{Stocznia Gdynia, Type 8234 containership,
general arrangement; drawing 8234/1-DZ/0110-001;
8234-series reference represented by \emph{MSC Florida}/\emph{Norasia Valparaiso}}. \\

\midrule

P01 &
Patrol &
\href{https://www.austal.com/sites/default/files/2025-12/Austal%20Offshore%20Patrol%2060%20-%20Maritime%20Security%20variant%20datasheet.pdf}
{Austal, \emph{Offshore Patrol 60}, Maritime Security variant,
vessel datasheet and general arrangement}. \\

P02 &
Patrol &
\href{https://www.austal.com/sites/default/files/2025-12/Austal%20Patrol%2040%20-%20Guardian-class%20datasheet.pdf}
{Austal, \emph{Patrol 40}, Guardian-class patrol vessel,
vessel datasheet and general arrangement}. \\

P03 &
Patrol &
\href{https://www.austal.com/sites/default/files/2025-12/Austal%20Patrol%2058%20Cape-Class%20%28Original%29%20datasheet.pdf}
{Austal, \emph{Patrol 58}, Cape-class patrol vessel,
vessel datasheet and general arrangement}. \\

\bottomrule
\caption{Source provenance of the 23 reconstructed real-reference arrangements.
Public links identify the vessel and source documentation used for reconstruction.
Drawing numbers, yard numbers, revisions, and dates are reported where available.
For sources for which the original hosting page could not be recovered, the
available document provenance is reported directly.}
\label{tab:app_ga_sources}
\end{longtable}
}

\section{Stability calibration and component variation}
\label{app:gm_calibration}

The stability model combines arrangement-dependent functional masses with a permanent mass contribution representing vessel mass not explicitly resolved by the compartment model.
Family-specific reference loading conditions are based on the design-draught-to-depth ratios ($T/D$) measured from the available real-reference arrangements, which are also used later as diagnostic comparison cases.

The remaining family-level parameters are determined from generated layouts.
An initial calibration set of 30 layouts per vessel family (180 layouts in total) is used to estimate $s_f$, $LCG_{\mathrm{perm}}$, and $KG_{\mathrm{perm}}$.
Once fixed, these parameters are applied to subsequent candidate arrangements without further adjustment.

For each vessel family $f$, the representative reference draught ratio $(T/D)_{\mathrm{ref},f}$ is obtained from the real-reference $T/D$ data.
For a represented hull, the corresponding reference draught is
\begin{equation}
T_{\mathrm{ref},f}
=
(T/D)_{\mathrm{ref},f}\,D ,
\end{equation}
and hydrostatic integration at this draught provides the reference displacement $\Delta_{\mathrm{ref}}$.
This displacement defines the total vessel mass used by the stability model.

For functional region $i$, fuel and ballast are assigned directly from their physical liquid densities, while the remaining functional classes use the relative mass-intensity coefficients defined in the main paper.
The represented mass is therefore
\begin{equation}
m_i =
\begin{cases}
0.98\,\rho_{y_i}V_i,
& y_i \in \{\mathrm{fuel},\mathrm{ballast}\}, \\[3pt]
s_f\,\tilde{\rho}_{y_i}V_i,
& \text{otherwise},
\end{cases}
\end{equation}
where $\rho_{y_i}$ is the physical liquid density and $\tilde{\rho}_{y_i}$ is the proxy mass-intensity coefficient. 
The parameter ($s_f$) scales the non-liquid functional masses relative to the reference displacement while preserving their relative values.
Fuel and ballast tanks are evaluated at $98\%$ fill, while the remaining functional spaces are evaluated at full capacity.

The directly represented liquid mass is
\begin{equation}
M_{\mathrm{liq}}
=
\sum_{i\in\mathcal{L}}
0.98\,\rho_{y_i} V_i ,
\end{equation}
where $\mathcal{L}=\{\mathrm{fuel},\mathrm{ballast}\}$.
The unscaled non-liquid relative mass is
\begin{equation}
\widetilde{M}_{\mathrm{nonliq}}
=
\sum_{i\notin\mathcal{L}}
\tilde{\rho}_{y_i} V_i ,
\end{equation}
such that the represented compartment mass becomes
\begin{equation}
M_{\mathrm{compartment}}
=
M_{\mathrm{liq}}
+
s_f\widetilde{M}_{\mathrm{nonliq}} .
\end{equation}
The non-liquid coefficients prescribe relative mass relationships.
Their absolute contribution is controlled by $s_f$, while the physical fuel and ballast masses remain unscaled.

Because detailed component masses are unavailable for the reconstructed arrangements, the fraction of total vessel mass represented directly by the functional arrangement cannot be identified from the available data.
Early-stage ship design commonly estimates unresolved weight groups and their centres using empirical or parametric information from similar vessels~\cite{SchneekluthShipDesignEfficiency1998}.
The present model therefore introduces a represented-mass fraction $\eta$ as a closure parameter,
\begin{equation}
M_{\mathrm{compartment}}
\approx
\eta\Delta_{\mathrm{ref}} .
\end{equation}
For this study, $\eta=0.50$ is adopted as a fixed conceptual-design assumption. 
For the present study, $\eta=0.50$ is adopted as a fixed conceptual-design calibration assumption.
It should not be interpreted as a standard vessel weight fraction and should be replaced when more detailed vessel-mass information becomes available.



Using the 30 calibration layouts for each vessel family, a single
family-specific scale $s_f$ is determined by minimising the squared
difference between the represented compartment mass and the target
fraction $\eta\Delta_{\mathrm{ref}}$ across the calibration set:
\begin{equation}
s_f
=
\underset{s}{\arg\min}
\sum_{k\in\mathcal{C}_f}
\left[
M_{\mathrm{liq},k}
+
s\widetilde{M}_{\mathrm{nonliq},k}
-
\eta\Delta_{\mathrm{ref},k}
\right]^2 ,
\label{eq:family_mass_scale}
\end{equation}
where $\mathcal{C}_f$ denotes the family calibration set.
The resulting scale preserves the prescribed relative mass relationships between non-liquid functions while setting their overall magnitude relative to the reference displacement.

The remaining vessel mass is represented by
\begin{equation}
M_{\mathrm{perm}}
=
\Delta_{\mathrm{ref}}
-
M_{\mathrm{compartment}},
\end{equation}
such that
\begin{equation}
M_{\mathrm{total}}
=
M_{\mathrm{compartment}}
+
M_{\mathrm{perm}}
=
\Delta_{\mathrm{ref}} .
\end{equation}
In the implementation, this unresolved mass is stored as lightship and residual contributions with a common mass centre.
They are combined here because only their summed mass and common moment enter the equilibrium calculation.

The permanent contribution therefore represents unresolved hull structure, outfit, and other vessel mass that cannot be determined from the functional arrangement alone.
Its longitudinal and vertical centres are calibrated at family level.
For an arrangement, the combined mass moments~\cite{papanikolaouShipDesignMethodologies2014} give,

{\scriptsize
\begin{equation}
LCG
=
\frac{
\displaystyle \sum_i (m_i x_i)
+
M_{\mathrm{perm}}LCG_{\mathrm{perm}}
}{
M_{total}
},
\qquad
KG
=
\frac{
\displaystyle \sum_i (m_i z_i)
+
M_{\mathrm{perm}}KG_{\mathrm{perm}}
}{
M_{total}
},
\label{eq:combined_mass_centres}
\end{equation}}

where $x_i$ and $z_i$ are the longitudinal and vertical centroids of functional region $i$.
For each vessel family, $LCG_{\mathrm{perm}}$ is adjusted iteratively until the
mean $LCG$ across the 30 calibration arrangements matches the corresponding
mean hull $LCB$.

For each calibration arrangement, the hull hydrostatics are then solved for equilibrium draught and trim~\cite{papanikolaouShipDesignMethodologies2014}.
At equilibrium,
\begin{equation}
\Delta(T_{\mathrm{eq}})
=
M_{\mathrm{total}},
\qquad
LCB=LCG .
\end{equation}
The hydrostatic properties at the solved condition provide $KB$ and $BM$, with
\begin{equation}
BM
=
\frac{I_T}{\nabla},
\end{equation}
where $I_T$ is the transverse second moment of the waterplane and $\nabla$ is the displaced volume.

For the prescribed fuel and ballast loading condition, a free-surface correction is applied to each partially filled liquid region.
Because detailed tank geometry is not available, the free surface of each liquid region $j$ is approximated as rectangular using its longitudinal length ($l_j$) and transverse breadth ($b_j$), giving
\begin{equation}
I_{\mathrm{FS},j}
=
\frac{l_j b_j^3}{12}.
\end{equation}
The total free-surface correction~\cite{papanikolaouShipDesignMethodologies2014} is then
\begin{equation}
\Delta GM_{\mathrm{FSM}}
=
\frac{
\displaystyle
\sum_{j\in\mathcal{T}}
\rho_j I_{\mathrm{FS},j}
}{
\Delta_{\mathrm{ref}}
},
\end{equation}
where $\mathcal{T}$ contains the partially filled fuel and ballast regions and $\rho_j$ is the corresponding liquid density.

After the equilibrium hydrostatics and free-surface correction are known, a target corrected metacentric height corresponds to
\begin{equation}
KG_{\mathrm{target}}
=
KB+BM-\Delta GM_{\mathrm{FSM}}-GM_{\mathrm{target}}.
\label{eq:kg_target}
\end{equation}
The family-level $KG_{\mathrm{perm}}$ is calibrated iteratively through Eq.~\eqref{eq:combined_mass_centres} so that the mean corrected-$GM$ distribution across the calibration arrangements is centred near $GM_{\mathrm{target}}=2.5$~m.
The $2.5$~m value is an early-design calibration target rather than a regulatory or family-specific stability requirement.

\begin{table}[t]
\centering
\scriptsize
\begin{tabular}{lcccc}
\toprule
Family &
$(T/D)_{\mathrm{ref}}$ &
$s_f$ &
$(LCG/L)_{\mathrm{perm}}$ &
$(KG/D)_{\mathrm{perm}}$ \\
\midrule
Yacht     & 0.64 & 0.125 & 0.433 & 0.200 \\
OSV       & 0.66 & 0.095 & 0.472 & 0.481 \\
Bulker    & 0.76 & 0.197 & 0.527 & 0.562 \\
Tanker    & 0.77 & 0.170 & 0.528 & 0.320 \\
Container & 0.74 & 0.168 & 0.519 & 0.496 \\
Patrol    & 0.64 & 0.097 & 0.417 & 0.096 \\
\bottomrule
\end{tabular}
\caption{Family-level parameters used by the calibrated stability model.
The reference draught ratios are based on the design draught and depth of the available real-reference arrangements.
The non-liquid mass scales and permanent mass centres are obtained from the family-level calibration.}
\label{tab:weight_calibration}
\vspace{-10pt}
\end{table}

A further 100 layouts per family, extracted from the source 50\,000 dataset, are then used as a calibration-check set to evaluate the resulting distributions and their variation.
The resulting parameters are shown in Table~\ref{tab:weight_calibration}.
The arrangement-specific compartment volumes and centroids determine the functional mass distribution, after which the same equilibrium and free-surface calculations are applied.
The corrected metacentric height is then
\begin{equation}
GM_{\mathrm{corr}}
=
KB+BM-KG-\Delta GM_{\mathrm{FSM}}.
\label{eq:sup_gm_components}
\end{equation}

\begin{figure}[t]
\centering
\includegraphics[width=0.95\linewidth]{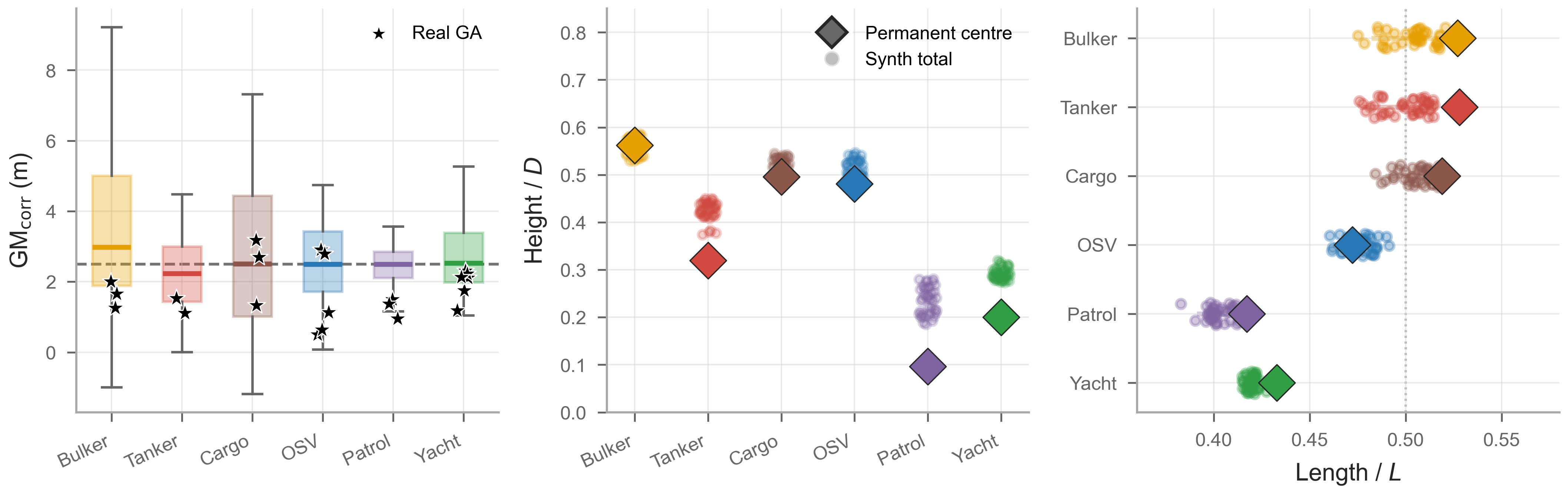}
\caption{Calibration evidence for the stability model.
The left panel compares the corrected-$GM$ distributions obtained across the 100-layout calibration-check set for each vessel family with diagnostic values obtained by applying the same model to the available real-reference arrangements.
The dashed line indicates the $2.5$~m calibration target.
The centre panel compares the calibrated permanent vertical centre with the resulting arrangement-dependent total $KG/D$ distributions.
The right panel compares the calibrated permanent longitudinal centre with the resulting arrangement-dependent total $LCG/L$ distributions.}
\label{fig:sup_gm_calibration}
\vspace{-10pt}
\end{figure}

Figure~\ref{fig:sup_gm_calibration} shows the resulting behaviour after the family-level calibration is fixed.
The spread within each family arises from differences in the generated mass distributions, solved hydrostatics, and free-surface corrections.

The corrected stability can be separated into its principal components by defining
\begin{equation}
KM
=
KB+BM ,
\end{equation}
so that Equation~\ref{eq:sup_gm_components} becomes
\begin{equation}
GM_{\mathrm{corr}}
=
KM-KG-\Delta GM_{\mathrm{FSM}} .
\end{equation}

\begin{figure}[t]
\centering
\includegraphics[trim={0cm 7.5cm 0cm 0cm}, clip, width=0.95\linewidth]
{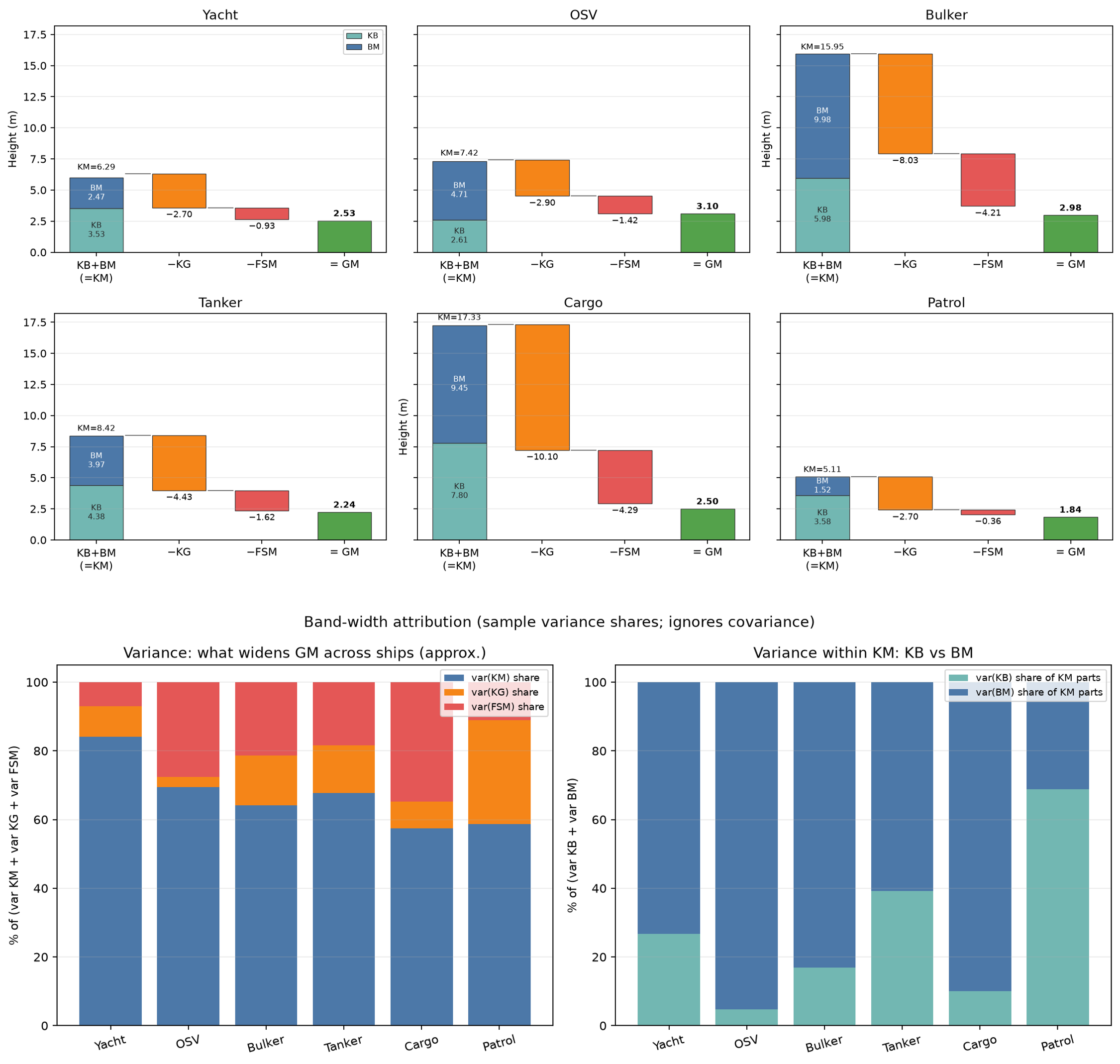}
\caption{Component accounting of corrected metacentric height for the six vessel families.
The hydrostatic contribution $KM=KB+BM$ is separated into $KB$ and $BM$, followed by subtraction of the vertical centre of gravity $KG$ and the free-surface correction $\Delta GM_{\mathrm{FSM}}$ to obtain $GM_{\mathrm{corr}}$.}
\label{fig:sup_gm_components}
\vspace{-5pt}
\end{figure}

\subsection{Variation of corrected-GM components}

Figure~\ref{fig:sup_gm_components} shows the relative magnitude of the hydrostatic, mass-distribution, and free-surface terms.
Although the family-level calibration is fixed, these components vary between arrangements.

To identify which terms contribute most strongly to this variation, sample variances are calculated across the 100 calibration-check arrangements for each vessel family.
Covariance-free variance shares are defined as
\begin{equation}
s_X
=
\frac{\operatorname{Var}(X)}
{\operatorname{Var}(KM)
+\operatorname{Var}(KG)
+\operatorname{Var}(\Delta GM_{\mathrm{FSM}})},
\end{equation}
for
\begin{equation}
X\in
\left\{
KM,\,
KG,\,
\Delta GM_{\mathrm{FSM}}
\right\}.
\end{equation}
These shares compare the relative variation of the three components rather than providing an exact decomposition of $\operatorname{Var}(GM_{\mathrm{corr}})$, because covariance between the terms is omitted.

The hydrostatic contribution is further separated using
\begin{equation}
s_Y^{KM}
=
\frac{\operatorname{Var}(Y)}
{\operatorname{Var}(KB)+\operatorname{Var}(BM)},
\qquad
Y\in\{KB,BM\},
\end{equation}
which identifies whether $KB$ or $BM$ contributes more strongly to the variation in $KM$.

\begin{figure}[t]
\centering
\includegraphics[trim={0cm 0cm 0cm 10cm}, clip, width=0.95\linewidth]
{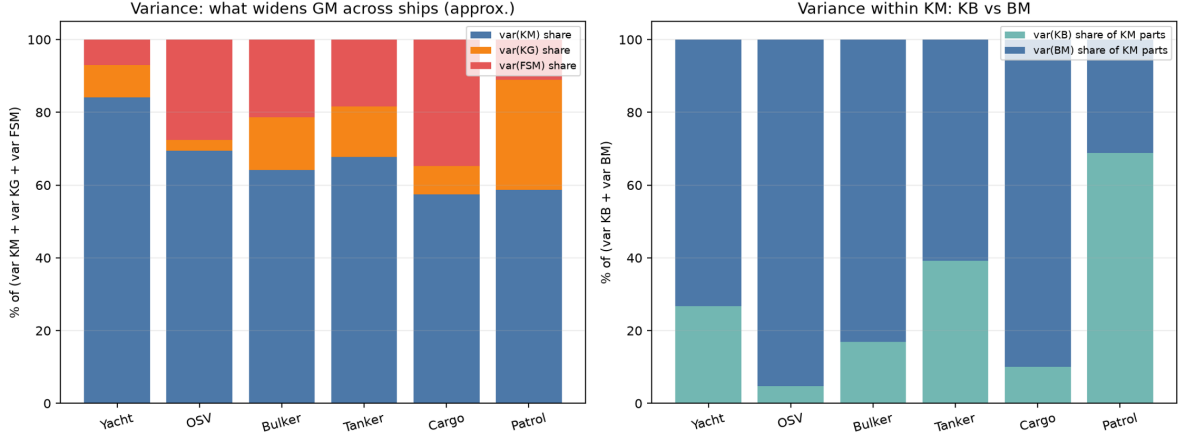}
\caption{Variance attribution of the corrected-stability components across the 100 calibration-check arrangements for each vessel family.
The left panel gives covariance-free sample-variance shares for $KM$, $KG$, and $\Delta GM_{\mathrm{FSM}}$.
The right panel further separates variation in $KM=KB+BM$ into contributions from $KB$ and $BM$.
}
\label{fig:sup_gm_decomposition}
\vspace{-5pt}
\end{figure}

Figure~\ref{fig:sup_gm_decomposition} shows that $KM$ provides the largest contribution to the corrected-$GM$ spread across the six vessel families.
For most families, this hydrostatic variation is dominated by $BM$, while the patrol vessel shows a stronger contribution from $KB$.
The remaining sources are family-dependent.
Free-surface variation is more important for the OSV and container vessel, while $KG$ contributes more strongly for the patrol vessel.
The corrected-$GM$ spread therefore reflects a combination of hydrostatic, mass-distribution, and free-surface effects.

\section{Fixed-envelope reconstruction visualisations}

\begin{figure}[h]
\centering
\begin{subfigure}[t]{\linewidth}
    \centering
    \includegraphics[width=0.95\linewidth]{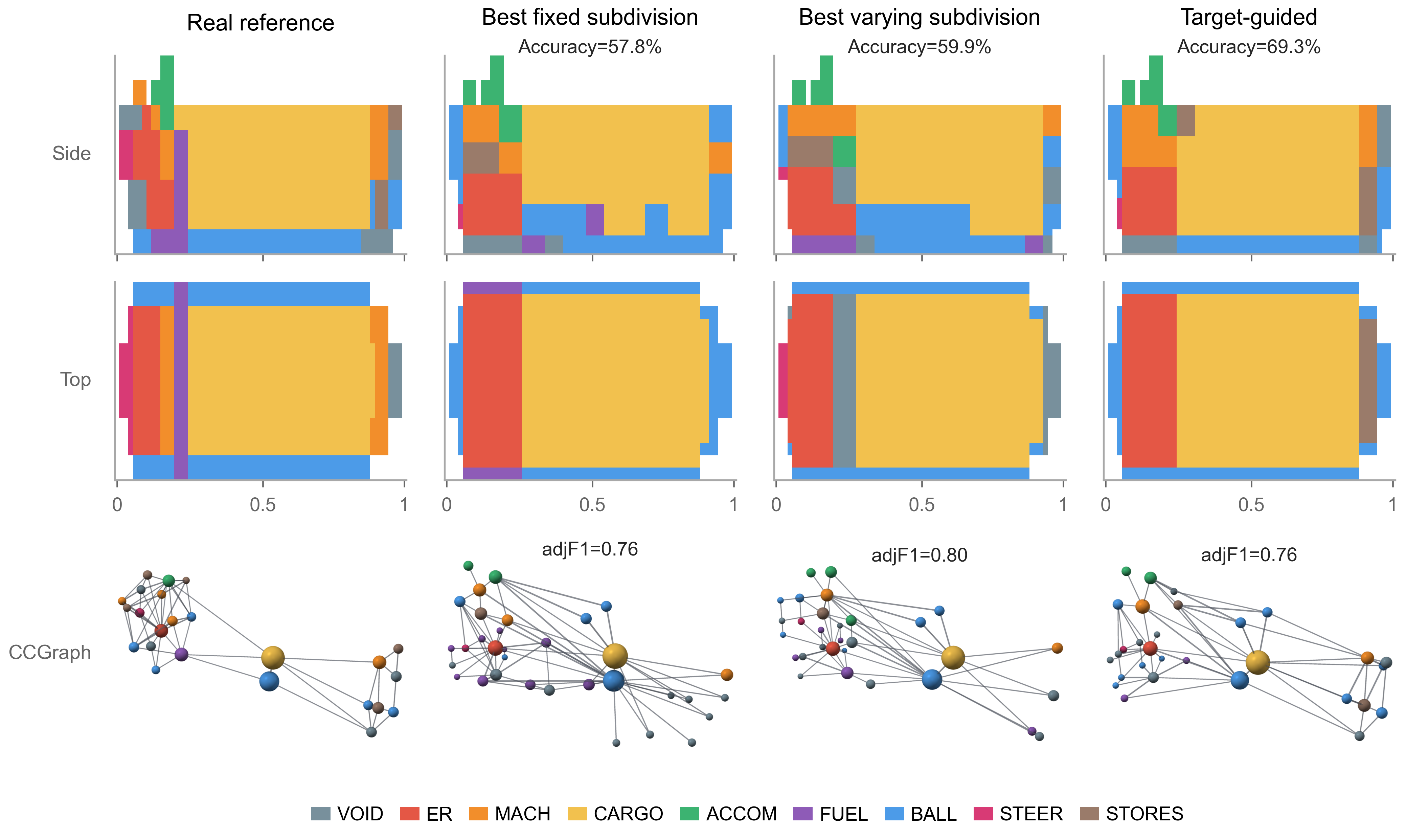}
    \caption{Tanker reference and accuracy-selected reconstructions.
    Accuracy increases from 57.8\% on the fixed subdivision to 59.9\% when subdivision and assignment both vary.
    Target guidance increases the reconstruction to 69.3\%.}
    \label{fig:e2_tanker}
\end{subfigure}

\vspace{5pt}

\begin{subfigure}[t]{\linewidth}
    \centering
    \includegraphics[width=0.95\linewidth]{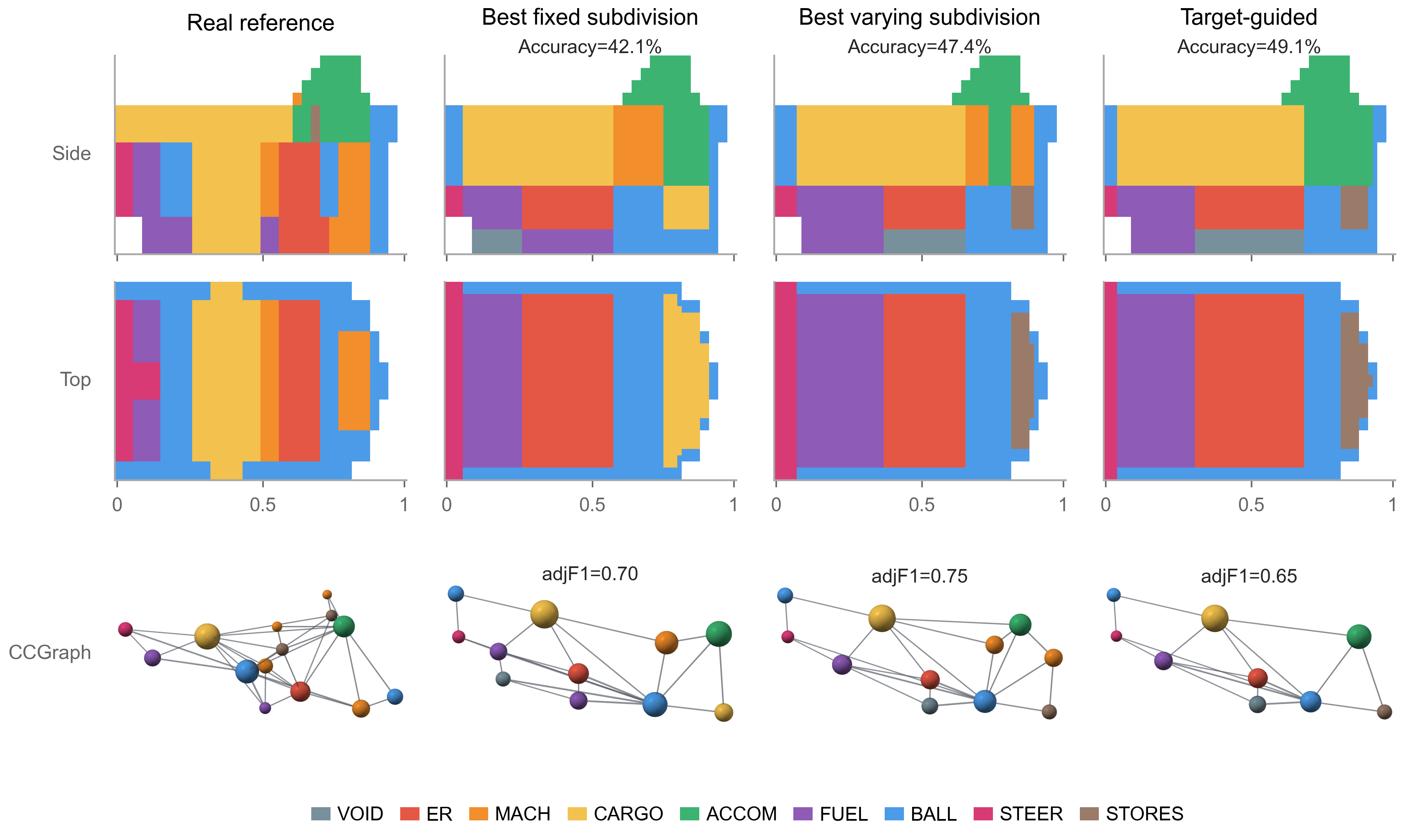}
    \caption{OSV reference and accuracy-selected reconstructions.
    Accuracy increases from 42.1\% on the fixed subdivision to 47.4\% with subdivision variation, followed by a smaller target-guided increase to 49.1\%.}
    \label{fig:e2_osv}
\end{subfigure}

\caption{Fixed-envelope reconstruction examples.
Each row compares the real reference with the best fixed-subdivision candidate, the best varying-subdivision candidate, and the target-guided diagnostic.
The upper view shows the longitudinal arrangement and the lower view shows the selected horizontal deck.
The target-guided result uses the real reference labels and is therefore a diagnostic rather than ordinary generator performance.}
\label{fig:e2_examples}
\end{figure}

Experiment~2 showed that reconstruction quality generally improves when the subdivision is allowed to vary in addition to the compartment assignment, although the magnitude of this improvement differs between vessel families.
The target-guided diagnostic further indicates how much of the remaining mismatch can be reduced through more favourable compartment assignment within a valid subdivision.
Figure~\ref{fig:e2_examples} provides explicit tanker and OSV examples of this progression, allowing the spatial differences between the reference, fixed-subdivision reconstruction, varying-subdivision reconstruction, and target-guided result to be compared directly.

\end{document}